\documentclass[aps,showpacs,prb,twocolumn,superscriptaddress]{revtex4-2}
\makeatletter

\newcommand{\Rmnum}[1]{\expandafter\@slowromancap\romannumeral #1@}
\makeatother
\usepackage[english]{babel}
\usepackage{amsmath}
\usepackage{graphicx}
\usepackage{subfigure}
\usepackage{color}\usepackage[colorlinks=true,citecolor=blue,urlcolor=blue,linkcolor=blue,hyperindex]{hyperref}
\usepackage{amsfonts}
\usepackage{tikz}
\usepackage{esint}
\usepackage{mathrsfs}
\usepackage{verbatim}
\usepackage[normalem]{ulem}
\usepackage{titlesec}
\usepackage{appendix}
 \usepackage{multirow}
\newcommand{\bea}{\begin{eqnarray}}
\newcommand{\eea}{\end{eqnarray}}

\begin{document}
\title{Studying line defect at Deconfined Quantum Criticality via fuzzy sphere regularization}
\author{Shutao Liu}
\affiliation{Department of Physics and State Key Laboratory of Surface Physics, Fudan University, Shanghai 200433, P.R. China}
\author{Shuai Yang}	
\affiliation{Department of Physics and State Key Laboratory of Surface Physics, Fudan University, Shanghai 200433, P.R. China}
\affiliation{Department of Physics, Hong Kong University of Science and Technology, Clear Water Bay, Hong Kong, China}
\author{Jie Lou}
\author{Yan Chen}
\email{yanchen99@fudan.edu.cn}
\affiliation{Department of Physics and State Key Laboratory of Surface Physics, Fudan University, Shanghai 200433, P.R. China}
\affiliation{Shanghai Branch, Hefei National Laboratory, Shanghai 201315, P.R. China}

\date{\today}
\begin{abstract}
The interplay between bulk critical fluctuations and nontrivial topology can enrich defect physics and give rise to novel defect universality classes. Understanding the fate of such defects therefore constitutes an important open problem. In this work, we studied a particularly simple setting: a (0+1)-dimensional pinning-field defect coupled to a (2+1)-dimensional deconfined quantum critical bulk. Using the fuzzy-sphere regularization, we numerically investigated the defect operator spectrum and extracted several universal quantities characterizing the defect conformal fixed point, including the scaling dimensions of defect-changing(creating) operators and the defect \(g\)-function. These results establish the first numerical characterization of line-defect conformal data at deconfined quantum criticality and may stimulate further investigations of defect critical phenomena in topological quantum critical matter.

\end{abstract}

\maketitle

\section{Introduction}
Critical phenomena in the presence of defects provide an important extension of conventional bulk universality\cite{Diehl_1997,CARDY1984514,cardy2008boundaryconformalfieldtheory}. The introduction of a boundary or defect partially break the conformal symmetry to its subgroup and introduces additional conformal data, thereby giving rise to new renormalization-group(RG) fixed points\cite{Bill__2016}. Consequently, a single bulk universality class may support multiple distinct defect or boundary universality classes, each characterized by its own universal critical exponents. Understanding how defects modify quantum critical behavior has therefore become a central theme in the study of defect conformal field theories(dCFTs).

A parallel development in defect physics concerns symmetry-protected topological(SPT) phases\cite{Haldane_model_PhysRevLett.61.2015,SPT_PhysRevB.87.155114,SPT_chen2013symmetryprotectedtopologicalorders}. Although these phases are gapped in the bulk, they can host gapless edge modes on open manifolds. An intriguing question arises when topological properties and quantum criticality coexist: how do the gapless edge degrees of freedom due to topology interact with critical bulk fluctuations, and can this interplay generate novel defect universality classes beyond those encountered in either conventional critical systems or gapped topological phases alone? In (1+1)-dimensional systems, studies of gapless symmetry-protected topological phases have partially addressed these questions\cite{E_Berg_PhysRevB.91.235309,gSPTjiang2017symmetryprotectedtopologicalluttinger,gSPTPhysRevB.97.165114,gSPTPhysRevLett.120.057001,gSPTPhysRevX.7.041048,gSPTverresen2020topologyedgestatessurvive,igSPT_PhysRevB.104.075132}, whereas in (2+1)-dimensional systems, recent numerical studies have explored this question in several settings, including unconventional surface universality classes at the transition between Haldane Paramagnetic phase and antiferromagnetic phases described by the three-dimensional O(3) universality class\cite{longzhang_PhysRevLett.118.087201}, the boundary criticality of topological insulators and topological superconductors\cite{Shen_2025,ge2025boundarycriticalitytwodimensionalcorrelated,Ge_2026_TI,toldin2025extraordinarytransitionedgecorrelated,edge_dqcp_liu2025edgemodestopologicalmott}, as well as interacting Gross–Neveu–Yukawa conformal field theories\cite{Jiang_2025}.

In this work, we investigate this interplay in the setting of deconfined quantum critical points (DQCPs)\cite{dqcp_doi:10.1126/science.1091806,dqcp_PhysRevB.70.144407,dqcp_PhysRevX.7.031051}, which realize a direct continuous quantum phase transition beyond the conventional Landau paradigm. This gapless state is constrained by emergent ’t Hooft anomaly\cite{dqcp_anomaly_Metlitski_2018}, and can be viewed as an intrinsically gapless symmetry-protected topological state\cite{igSPT_PhysRevB.104.075132,igSPT_Yang_2026}. These topological characteristics coexist with strong critical fluctuations and therefore provide a natural arena for exploring unconventional defect physics. This issue has been extensively investigated in the context of novel boundary universality classes\cite{edge_dqcp_Ma_2022,edge_dqcp_liu2025edgemodestopologicalmott,edge_dqcp_myersonjain2024pristinepseudogappedboundariesdeconfined,edge_dqcp_cui2026extraordinaryboundarycorrelationsdeconfined}. In contrast, the physics of (0+1)-dimensional line defects remains largely unexplored. 

Motivated by these considerations, we study the universal properties of one-dimensional line defects embedded in a (2+1)-D DQCP and examine how the combined effects of topology and criticality determine their low-energy behavior. This simple line defect can be obtained by turning on a relevant perturbation spatially localized on a line $x^1=x^2=\cdots=x^{d-1}=0$,
\begin{equation}
    S_{\text{dCFT}}=S_{\text{dqcp}}+\vec{h}\cdot \int dx^0\vec{\phi}(x^0)
\end{equation}
usually, $\vec{\phi}(x^0)$ denotes the order parameter field in deconfined quantum phase transitions and its scaling dimension satisfies $\Delta_{\vec{\phi}}<1$. Physically, it describes the response of the system to a spatially localized magnetic field, leading to the definition of a pinning field. With this construction, we aim to address two central questions in the following: (i) what is the infrared fate of the simplest pinning-field defect coupled to a DQCP bulk? Does it become screened, such the system flows to a pure bulk without any defect, or does it flow to a non-trivial defect fixed point characterized by its own defect operator spectrum? (ii) Can a one-dimensional line defect immersed in a DQCP bulk undergo spontaneous symmetry breaking and support stable long-range order?

To address these questions, we employ the fuzzy-sphere regularization, a nonperturbative approach to conformal field theories that has been developed extensively in recent years\cite{ising_fuzzy_spherePhysRevX.13.021009}. The method exploits the state-operator correspondence directly, enabling the extraction of conformal data such as operator scaling dimensions, correlation functions\cite{han2023conformalfourpointcorrelators3d}, operator-product-expansion (OPE) coefficients\cite{ope_Hu_2023}, and monotonic quantities along renormalization-group flows\cite{hu2024entropicffunction3dising}. Also, the fuzzy-sphere approach has also been successfully generalized to defect conformal field theories\cite{dCFT_Sphere_Hu_2024,dCFT_Sphere_Zhou_2024,dCFT_Sphere_dedushenko2024isingbcftfuzzyhemisphere,dCFT_Sphere_cuomo2024impuritiescuspgeneraltheory,dCFT_Sphere_Zhou_2025,dCFT_Sphere_sarma2026fortuitousuniversalitybosekondoimpurities,dCFT_Sphere_feng2026studying3donsurface}, providing a powerful framework for studying the interplay between defects and strongly interacting critical bulk degrees of freedom.

It is worth emphasizing that numerical evidence\cite{dqcp_models_PhysRevLett.110.185701,dqcp_models_PhysRevX.5.041048,dqcp_modelsPhysRevLett.125.257204,dqcp_models_PhysRevB.99.195110,dqcp_models_Liu_2019,dqcp_models_PhysRevLett.133.100402,Zhou_2024} accumulated so far suggests that many lattice realizations of the DQCP are weakly first-order transitions rather than truly continuous ones. Nevertheless, these systems exhibit pronounced pseudocritical behavior over accessible length scales, possibly due to nearby genuine renormalization-group fixed points\cite{dqcp_pseudo_criticality_PhysRevB.102.020407,dqcp_pseudo_PhysRevB.102.201116}. From this perspective, the study of defects in the vicinity of the putative critical regime remains both meaningful and physically relevant, as it probes the universal physics governing systems close to a genuine deconfined critical point.

The remainder of this paper is organized as follows. In Sec.\ref{sec:defect_ham}, we introduce the bulk Hamiltonian formulated on the fuzzy sphere and describe how the line defect is incorporated into the model. In Sec.\ref{sec:defect_data}, we present the numerical results, including the defect conformal multiplet, correlation functions, OPE coefficients, and the defect $g$-function. Finally, in Sec.\ref{sec:summ_and_diss}, we summarize our main findings, discuss the limitations of the present work, and outline several promising directions for future research.

\section{Defect Hamiltonian}\label{sec:defect_ham}
We begin with an interacting fermionic model defined on the fuzzy sphere, which hosts a putative SO(5) deconfined quantum critical point. The bulk Hamiltonian is given by
\begin{equation}
\begin{split}
    H_{\text{dqcp}} = \int d\boldsymbol{r}_1d\boldsymbol{r}_2&\delta \left( \boldsymbol{r}_{1}-\boldsymbol{r}_2 \right)\Biggl[ 
     \boldsymbol{n}_0\left( \boldsymbol{r}_1 \right) \boldsymbol{n}_0\left( \boldsymbol{r}_2 \right) \\
    & \hphantom{u_0} -\sum_{i=1}^5{u_i\boldsymbol{n}_i\left( \boldsymbol{r}_1 \right) \boldsymbol{n}_i\left( \boldsymbol{r}_2 \right)} \Biggr],
\end{split}
\label{eq:ham_real}
\end{equation}
where the fermion bilinears $n^{i}(\boldsymbol{r})=\boldsymbol{\psi}^{\dagger}(\boldsymbol{r})\Gamma^{i}\boldsymbol{\psi}(\boldsymbol{r})$
denotes the five possible polarization channels associated with the Dirac matrices \(\Gamma^{i=1,\cdots,5}=\{\tau_x\otimes\mathbb{I},\tau_y\otimes\mathbb{I},\tau_z\otimes\vec{\sigma}\}\). The model possesses an exact SO(5) symmetry when the interaction strengths satisfy
$u_{1}=u_{2}=u_{3}=u_{4}=u_{5}$. One can also make an O(4) deformation by setting $u_5<u_{1,2,3,4}$ to support a potential O(4) symmetric deconfined point.

The bulk pseudocritical behavior of this model has been extensively investigated in previous works\cite{Zhou_2024,ys_o4dqcp_l6vw-6z79}. Here, we focus on the simplest realization of a line defect. Under the state-operator correspondence, a straight line defect in flat space is mapped to two localized imputities insertions at the north and south poles of the sphere. Thus, we introduce (0+1)-dimensional spatially localized magnetic fields(pinning fields), \(h_{N}\) and \(h_{S}\), coupling to the SO(5) vector field at the north and south poles of the fuzzy sphere. Such a defect has been studied from various aspects\cite{defect_Vojta_2000,defect_Sachdev_2003,defect_Parisen_Toldin_2017,defect_Assaad_2013,defect_allais2014magneticdefectlinecritical,defect_Florens_2006,defect_Franchi_2022,defect_Cuomo_2022} and it usually flows to conformal fixed points in several bulk CFTs\cite{defect_Cuomo_2022,defect_barrat2025linedefectcorrelatorsfermionic}. Without loss of generality, we choose the pinning field to couple to the fifth component of the SO(5) vector order parameter. The full Hamiltonian in the presence of the line defect is therefore written as
\[
H_d=H_{\text{dqcp}}+h_Nn_5(\theta=0,\varphi)+h_Sn_5(\theta=\pi,\varphi)
\]
By choosing a sufficiently large defect coupling strength $h_d=h_N=h_S$, the system can be driven close to the nontrivial defect fixed point. Throughout this work, we take \(h_d=1000\) as a representative parameter point for extracting universal defect conformal data because numerically we found $h_d=1000$ to be sufficiently close to the infrared defect fixed point, as will be demonstrated below from the saturation of the defect operator spectrum. To further verify the universality of the extracted conformal data, we also implement the infinite coupling limit \(h_d\rightarrow\infty\) directly by freezing the spins on the \(m=\pm s\) orbitals. The details of the defect deformation for another commonly studied O(4) DQCP, together with numerical results, are presented in the Sec. \ref{appsec:defect_in_o4_dqcp} of the Appendix.

\section{Defect conformal data}\label{sec:defect_data}
In this section, we show numerical results extracted from the Hamiltonian's energy spectrum. We mainly employ exact diagonalization (ED)\cite{zhou2025fuzzifiedjuliapackage} for systems with up to \(N_m=10\) orbitals to obtain defect conformal data 
These calculations allow us to access various quantities characterizing the defect conformal fixed point, including the defect operator spectrum, correlation functions, OPE coefficients, and the defect \(g\)-function.

Now, we briefly describe the overall roadmap when we present numerical results in the following. We first instivigate the operator flows as we deviate from the bulk fixed point when increasing defect strength $h_d$. Then we use state-operator correspondence to resolve the defect spectrum according to the branch rules of SO(N) group. At these defect fixed points, we extract several universal defect confromal data to clarify the defect universality class. 
\subsection{Operator flows from bulk to defect fixed point}
\begin{figure}[!htbp] 
    \centering
    \includegraphics[width=0.48\textwidth]{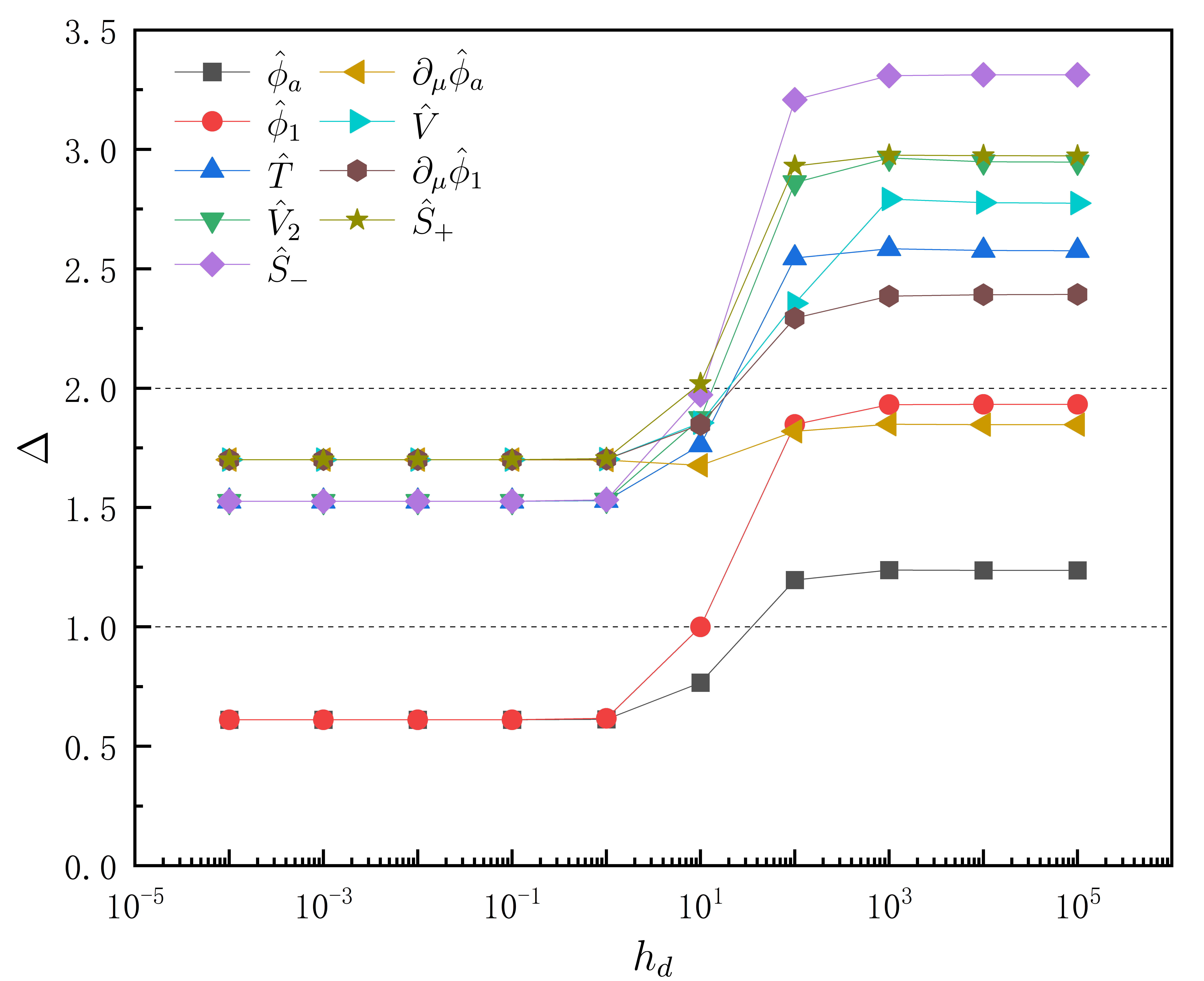} 
    \caption{Evolution of several low-lying excitation states along the RG flow from the bulk fixed point$(h_
d=0)$ to the defect fixed point$(h_
d\to \infty)$. For $h_d\ge100$, the extracted scaling dimensions approach stable values, indicating that the system is already close to the defect conformal fixed point. The results shown here are obtained for the bulk SO(5) deconfined quantum phase transition point\cite{Zhou_2024} on a system with size $N_o=6$.}
    \label{fig:on_dqcp_op_flow}
\end{figure}

To identify the low-energy operator content at the defect fixed point, we first examine how the low-lying spectrum evolves as the defect strength is increased. The evolution of the low-energy spectrum as a function of the defect strength \(h_d\) is shown in Fig.\ref{fig:on_dqcp_op_flow}. 
The splitting of degeneracies can be understood from the symmetry breaking induced by the defect (see, e.g.\cite{wiki:RepClassicalLieGroups}, the branching rules of SO($N$) representations). Specifically, the defect breaks (i) the spatial SO($3$) rotational symmetry, such that the states belonging to different \(L_z\) sectors that are degenerate in the bulk can split, and (ii) the internal flavor symmetry is reduced from SO(5) to SO(4). As a consequence, the bulk SO(5) vector operator \(\vec{\phi}\) decomposes into an SO($4$) vector \(\hat{\phi_a}(a=1,\ldots,4)\) together with an SO(4) singlet $\hat{\phi}_1$ corresponding to the component parallel to the pinning direction\footnote{To distinguish defect operators from bulk operators, we denote defect CFT operators by placing a $\hat{}$ over the corresponding symbols.}. Similarly, the descendant operator \(\partial_\mu\vec{\phi}\) decomposes into \(\partial_\mu\hat{\phi_a}\) and the derivative of the singlet component. Using the defect state-operator correspondence, we identify the lowest-energy SO($4$) vector $\hat{\phi}_a$ excitation with the tilt operator $\hat{t}$, while the lowest-energy SO($4$) singlet $\partial_\mu\hat{\phi}_1$ in the \(L_z=1\) sector is identified as the displacement operator $\hat{D}$. 

\begin{figure}[!htbp] 
    \centering
    \includegraphics[width=0.48\textwidth]{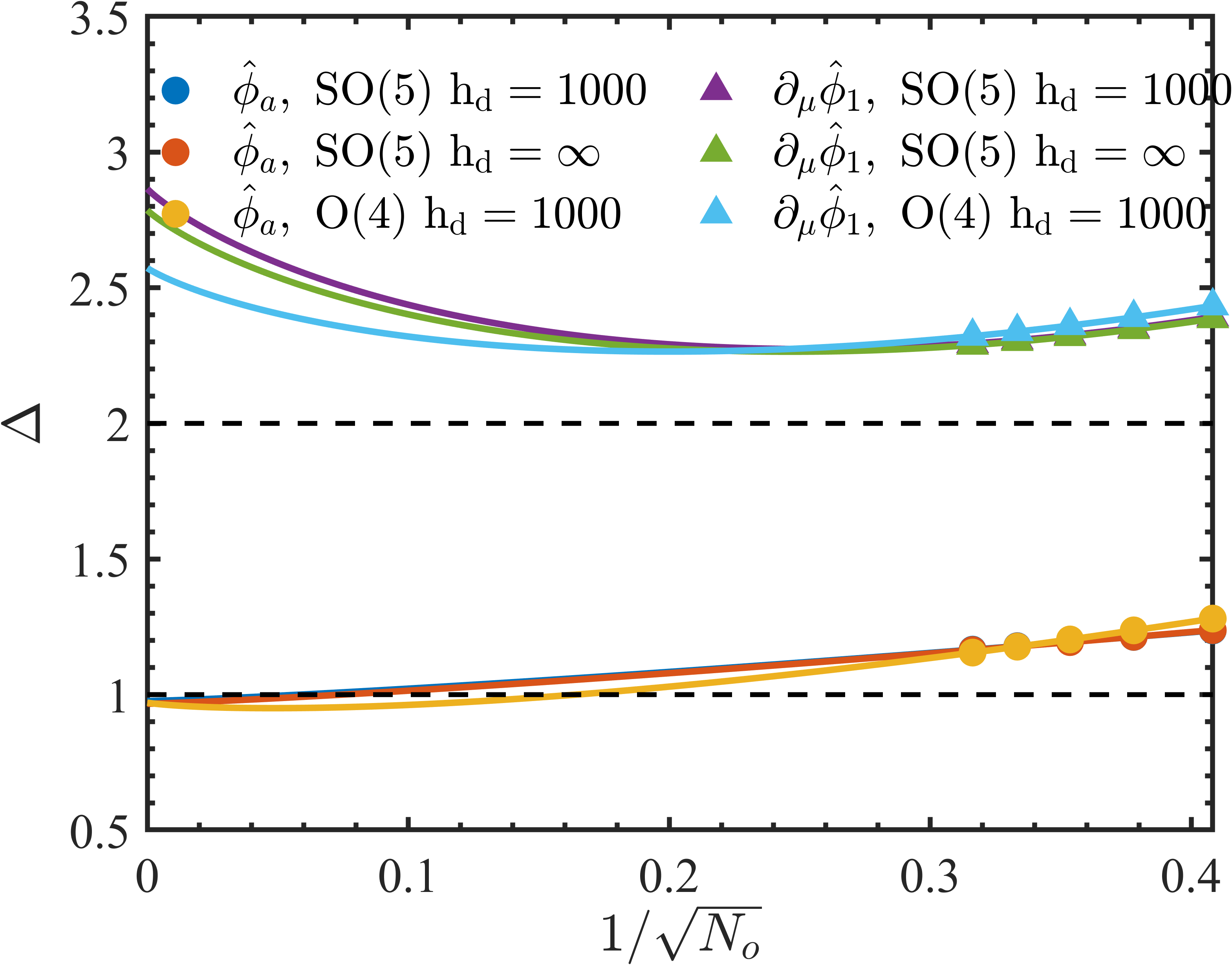} 
    \caption{Consistent finite size scaling results of operator $\hat{\phi_a}$ and $\partial_\mu\hat{\phi_1}$. Based on symmetry analysis, we identify these two operators as the tilt operator $\hat{t}$ and the displacement operator $\hat{D}$, respectively.}
    \label{fig:consist_scaling_tilt_disp}
\end{figure}

As shown in Fig.\ref{fig:consist_scaling_tilt_disp}, finite-size extrapolation of the scaling dimensions of $\hat \phi_a$ and $\partial_\mu\hat{\phi}_1$ shows that the former approaches the CFT prediction of 1. In contrast, $\partial_\mu\hat{\phi}_1$ exhibits considerably stronger finite-size effects. 
A more detailed discussion of this behavior is provided in the Discussion section(Sec.\ref{sec:summ_and_diss}). For these defect primary operators, we summarize the scaling dimensions extracted at finite system sizes in the 
Tab.\ref{tab:o5_wf_operator_dimensions}. 
Obviously, for the same SO(5) deconfined quantum critical bulk, the results obtained for different pinning-field strengths($h_d=1000\ \text{or}\ \infty$) are roughly consistent with one another. This consistency suggests that the observed behavior reflects the universal properties of line defects in the deconfined quantum critical bulk.

    \begin{table*}[htbp!]
  \centering
  \caption{Scaling dimensions of several primary operators in the defect CFT obtained from finite-size calculations. The data produced at different pinning field strengths $h_d=1000$ and $h_d=\infty$ are roughly consistent. The bulk Hamiltonian is tuned to the $SO(5)$ Deconfined quantum critical point. Here, $\text{Dim}$ and $C_2$ denote the dimension and the quadratic Casimir eigenvalue of the representations under the residual $SO(4)$ group.}
  
\begin{tabular}{c|ccc|ccc|ccc|c}
\hline\hline
\multirow{2}{*}{Operator} 
& \multirow{2}{*}{$L_z$} 
& \multirow{2}{*}{$\text{Dim}_{SO(4)}$} 
& \multirow{2}{*}{$C_2^{\mathrm{SO(4)}}$} 
& \multicolumn{3}{c|}{$h_d=1000$}
& \multicolumn{3}{c|}{$h_d=\infty$}
& \multirow{2}{*}{Notes} \\
\cline{5-10}
& & & 
& $N_o=10$ & $N_o=9$ & $N_o=8$ & $N_o=10$ & $N_o=9$& $N_o=8$ & \\
\hline
$\hat{\phi}_1$              
& 0 & 1 & 0 & 1.8314 & 1.8470& 1.8667& 1.8344
& 1.8495 & 1.8686 &  \\
$\hat{\phi}_a$              
& 0 & 4 & 3 & 2.2884 & 2.3021&2.3202 &2.2942 & 2.3077 & 2.3256 & tilt Op. \\
$\hat{S}_-$                 
& 0 & 1 & 0 & 3.1111 &3.1474& 3.1924& 3.1165& 3.1524 & 3.1972 &  \\
$\hat{V}_a$                 
& 0 & 4 & 3 & 2.8956 & 2.9057& 2.9072&2.8935 & 2.9008 & 2.8977 &  \\
$\hat{T}_{ab}$              
& 0 & 9 & 8 & 2.4462 & 2.4721& 2.5027& 2.4460& 2.4707 & 2.4997 &  \\
$\partial_\mu \hat{\phi}_1$ 
& 1 & 1 & 0 & 2.2884 & 2.3021& 2.3202&2.2942 & 2.3077 & 2.3256 & displacement Op. \\
$\partial_\mu \hat{\phi}_a$ 
& 1 & 4 & 3 & 1.7758 & 1.7893&1.8055 &1.7781 & 1.7910 & 1.8064 &  \\
\hline
$\hat{\phi}^{+-}$           
& 0 & 1 & 0 & 1.2569 & 1.2610& 1.2670&1.2762 & 1.2809 & 1.2877& defect changing Op. \\
$\hat{\phi}^{+0}$           
& 0 & 1 & 0 & 0.1648 &0.1663& 0.1682& 0.1671& 0.1687 & 0.1709 & defect creation Op. \\
\hline\hline
\end{tabular}
    \label{tab:o5_wf_operator_dimensions}

\end{table*}

\begin{figure*}[!htbp] 
    \centering
\includegraphics[width=0.16\textwidth]{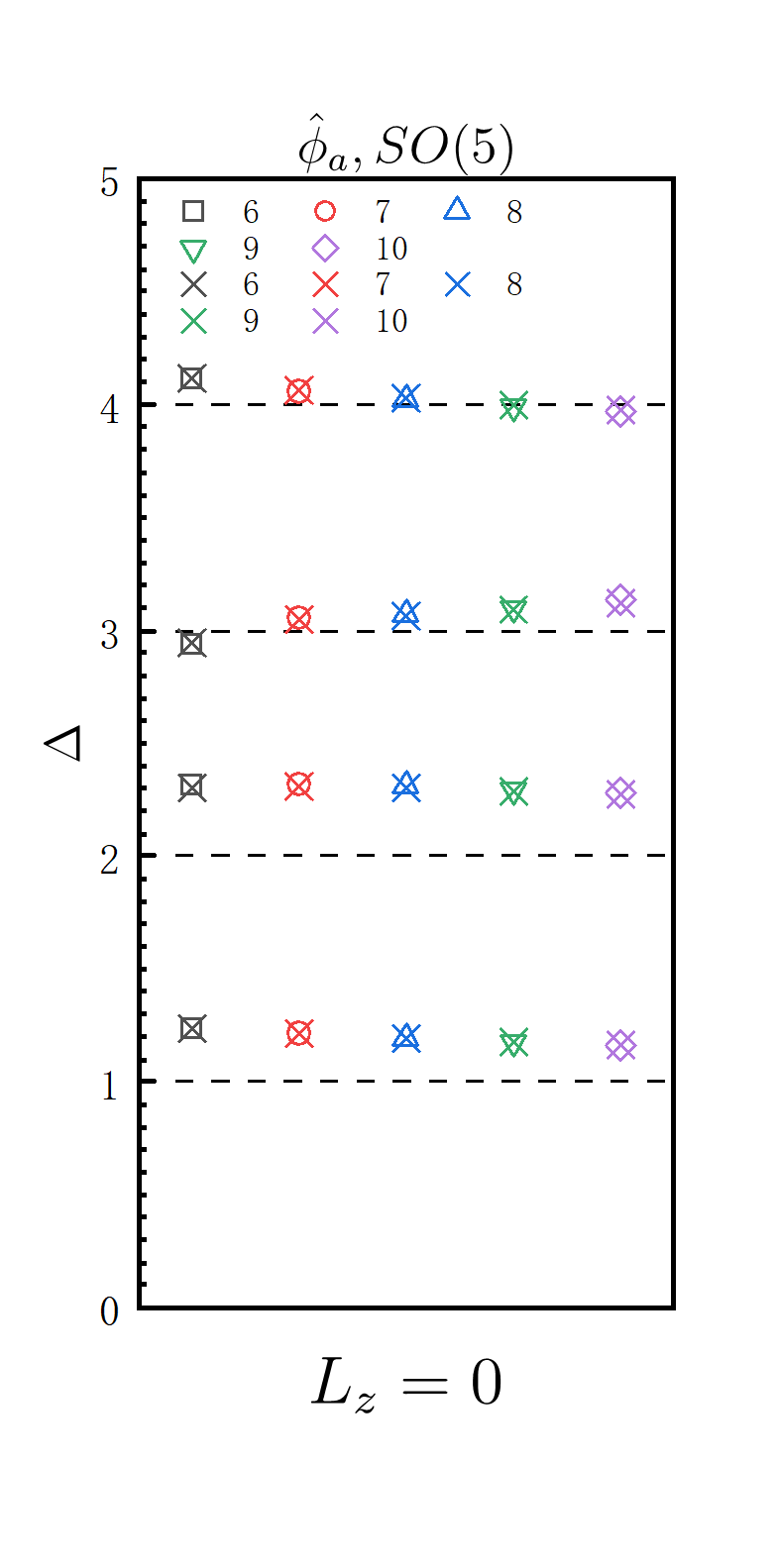}
\includegraphics[width=0.16\textwidth]{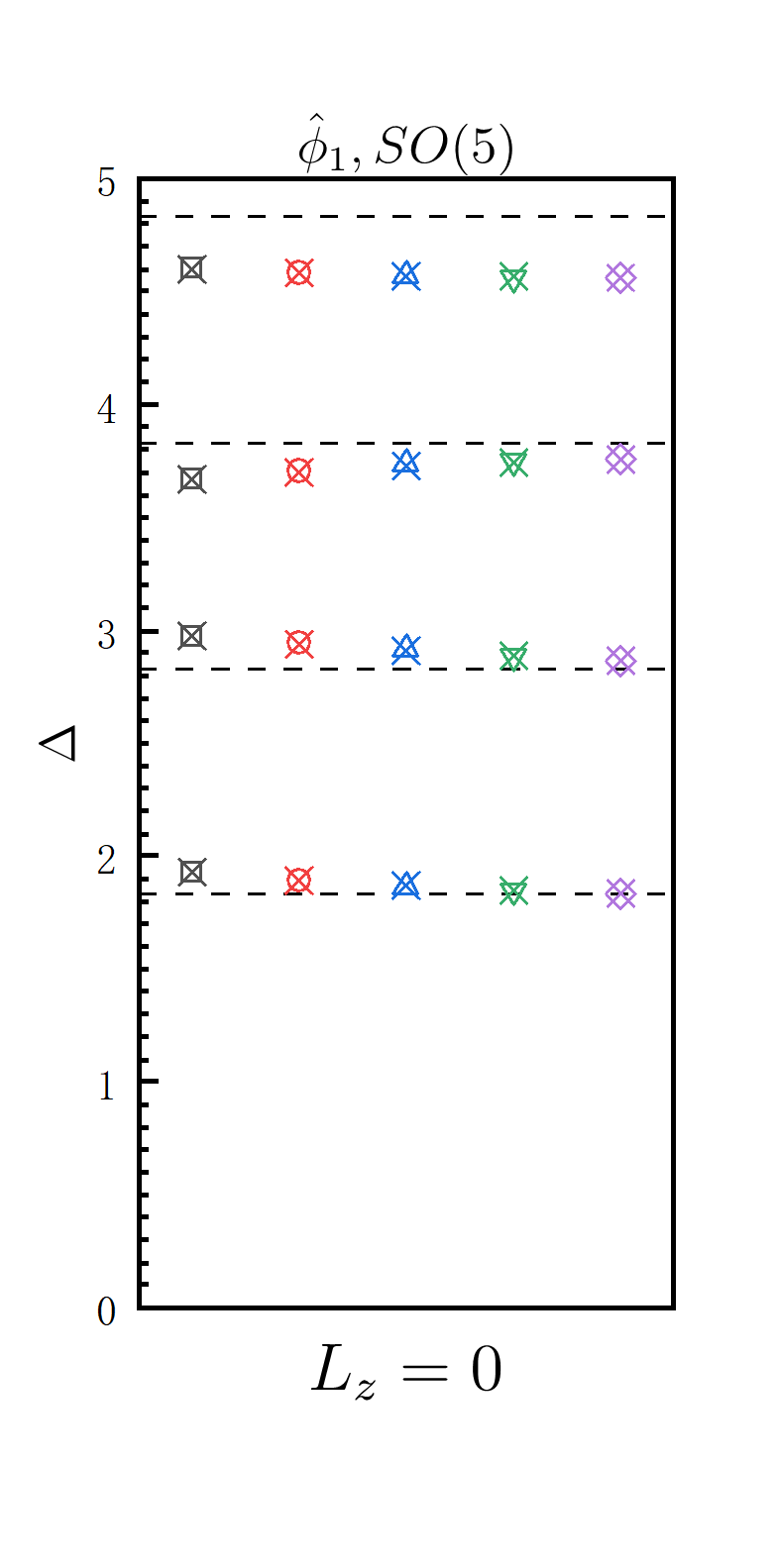}
\includegraphics[width=0.16\textwidth]{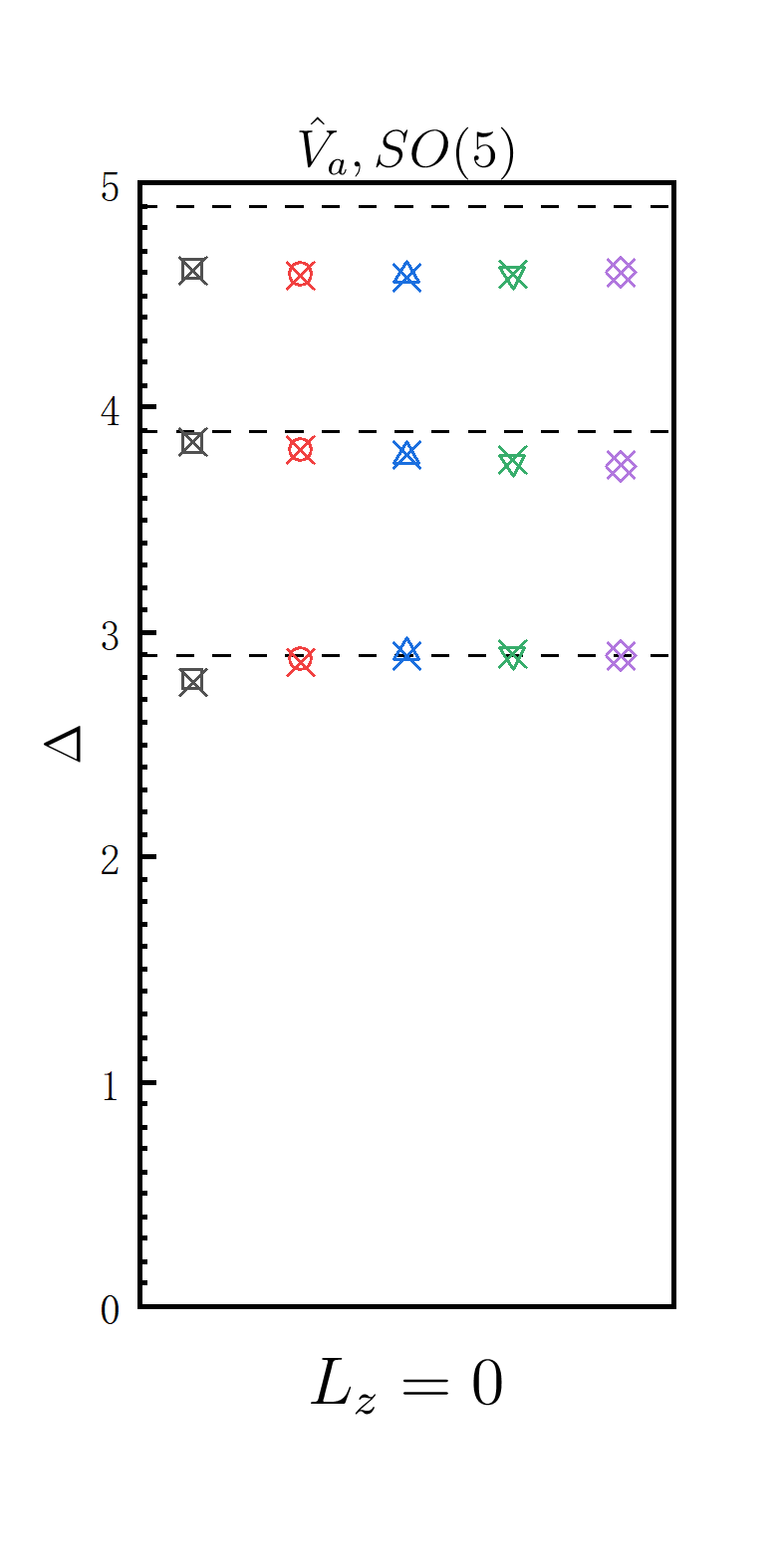}
\includegraphics[width=0.16\textwidth]{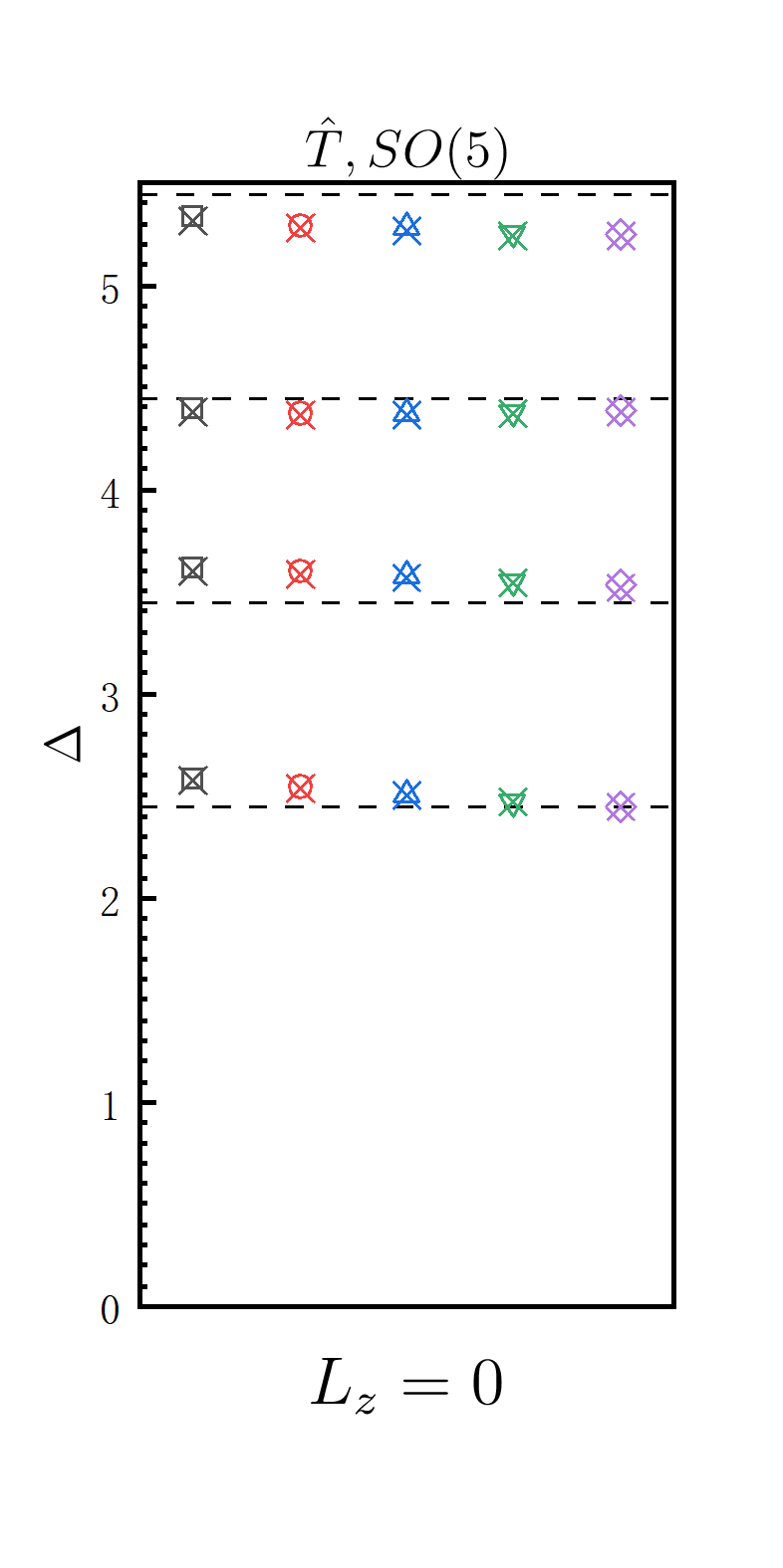}
\includegraphics[width=0.16\textwidth]{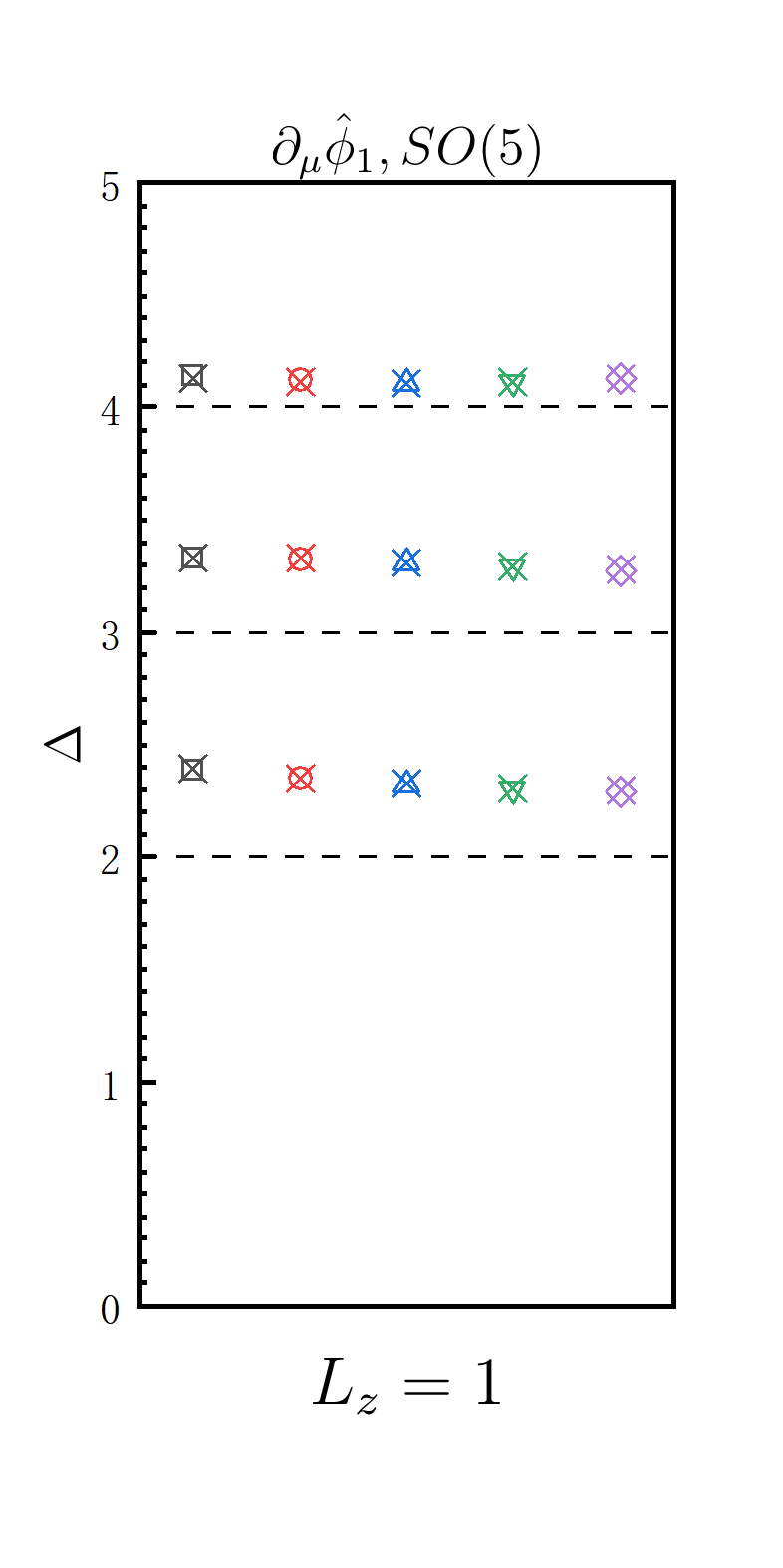}
\includegraphics[width=0.16\textwidth]{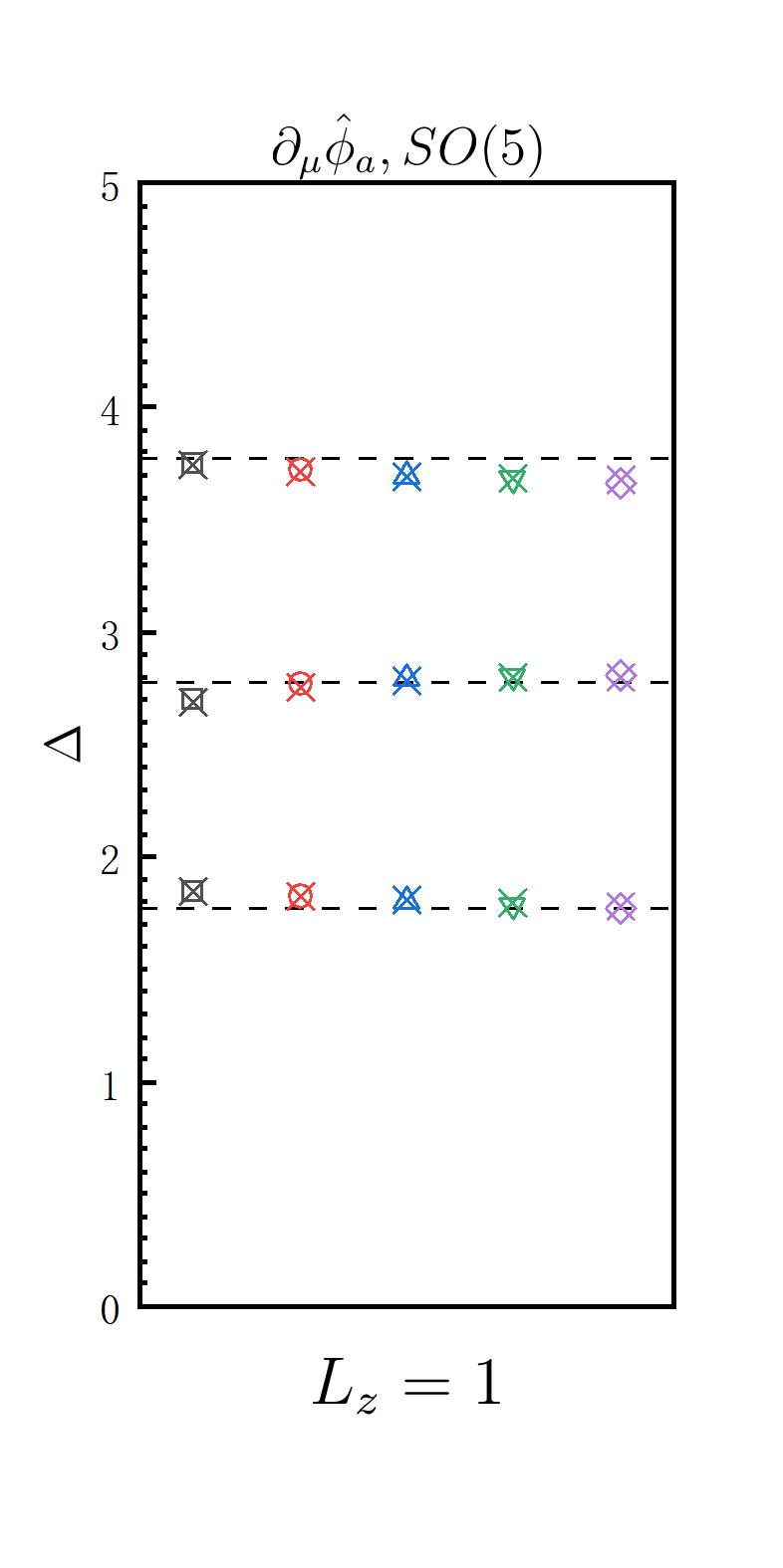}
    \caption{Line defect conformal multiplets, we present several lowest SO($4$) singlets $\hat\phi_1,\ \partial_\mu\phi_1$, vectors $\hat \phi_a,\ \hat V_a,\ \partial_\mu\phi_a$ and rank-2 tensor $\hat T_{ab}$. The horizontal gray lines indicate the operator scaling dimensions extracted from the largest system size accessible to ED, and are shown as reference values to improve the clarity of the figure. These results are produced at $h_d=1000$(open symbols) with subject to pseudo-critical SO(5) bulk theories. We also present the defect operator spectrum (cross symbols) obtained from an alternative realization of the $h_d=\infty$ pinning field, in which the spins are pinned in orbital space. Here, the bulk is tuned to be at SO(5) deconfined quantum critical point\cite{Zhou_2024}.
    }
    \label{fig:so5_real_multi}
\end{figure*}

Having identified the low-lying primary operators, we now demonstrate that these low-energy excitations can be reorganized into conformal multiplets as required by the defect conformal symmetry. This provides a straightforward consistency check that the infrared theory is governed by a defect CFT. Each tower consists of a primary operator together with its descendants generated by the action of the translation generator along the defect worldline. Since the translation generator carries scaling dimension one, descendant operators have scaling dimensions shifted by integers relative to the corresponding primary. As shown in Fig.\ref{fig:so5_real_multi}, the numerically obtained spectra exhibit the expected integer spacing and agree well with the defect conformal group, providing compelling evidence for the emergence of defect conformal symmetry.

\subsection{Correlation functions and OPE coefficients}
Build upon this defect fixed point, we now proceed to extract correlation functions and OPE coefficients. Together with the operator spectrum, these quantities constitute the essential conformal data that characterize a given defect CFT.

In particular, we focus on the one-point functions of bulk operators in the presence of the defect and the bulk-to-defect two-point correlation functions. Within the framework of radial quantization, conformal symmetry constrains these correlators to take the forms of
\begin{equation}
\begin{aligned}
        G_O&=\langle {O}(x)\rangle=
\frac{a_{ O}}
{|x_\perp|^{\Delta_{O}}}\\
G_{O\hat{O}}&=\langle {O}(x)\hat{O}(0)\rangle=
\frac{
b_{\hat{ O}}
}
{|x_\perp|^{\Delta_{O}-\Delta_{\hat{ O}}}
|x|^{2\Delta_{\hat{ O}}}},
\end{aligned}
\end{equation}
where \({O}\) denotes a bulk operator, \(\hat{{O}}\) is a defect operator. And, \(x_\perp=r\sin{\theta}\) denotes the perpendicular distance from the bulk point \(x=(r,\theta,\varphi)\) to the defect point. The coefficient \(a_{O}\) appearing in the one-point function and the bulk-defect OPE coefficient \(b_{\hat{ O}}\) are universal quantities that encode intrinsic information about the defect CFT. On the fuzzy sphere, these correlators can be evaluated using the state-operator correspondence under the Weyl transformation, hence, the one-point
and two-point functions can be written as inner products:
\begin{equation}
    \begin{aligned}
        G_{O}(x)&=\frac{\langle \hat{\mathbb{I}}|O(x)|\hat{\mathbb{I}}\rangle}{\langle 0|O(x)|0\rangle}\approx\frac{a_O}{(\sin{\theta})^{\Delta_O}}+O(1/R)\\
        G_{O\hat{O}}(x)&=\frac{\langle \hat{\mathbb{I}}|O(x)|\hat{{O}}\rangle}{\langle 0|O(x)|0\rangle}\approx\frac{b_{\hat{O}}}{(\sin{\theta})^{\Delta_O-\Delta_{\hat{O}}}}+O(1/R)
    \end{aligned}
    \label{eq:corr_reference}
\end{equation}
Here, $|\hat{\mathbb{I}}\rangle$ denotes the defect vacuum corresponding to the identity operator in the defect CFT, and $|\hat{O}\rangle$ is an excited state corresponds to the defect operator $\hat{O}$

\begin{figure*}[!htbp] 
    \centering
\includegraphics[width=0.24\textwidth]{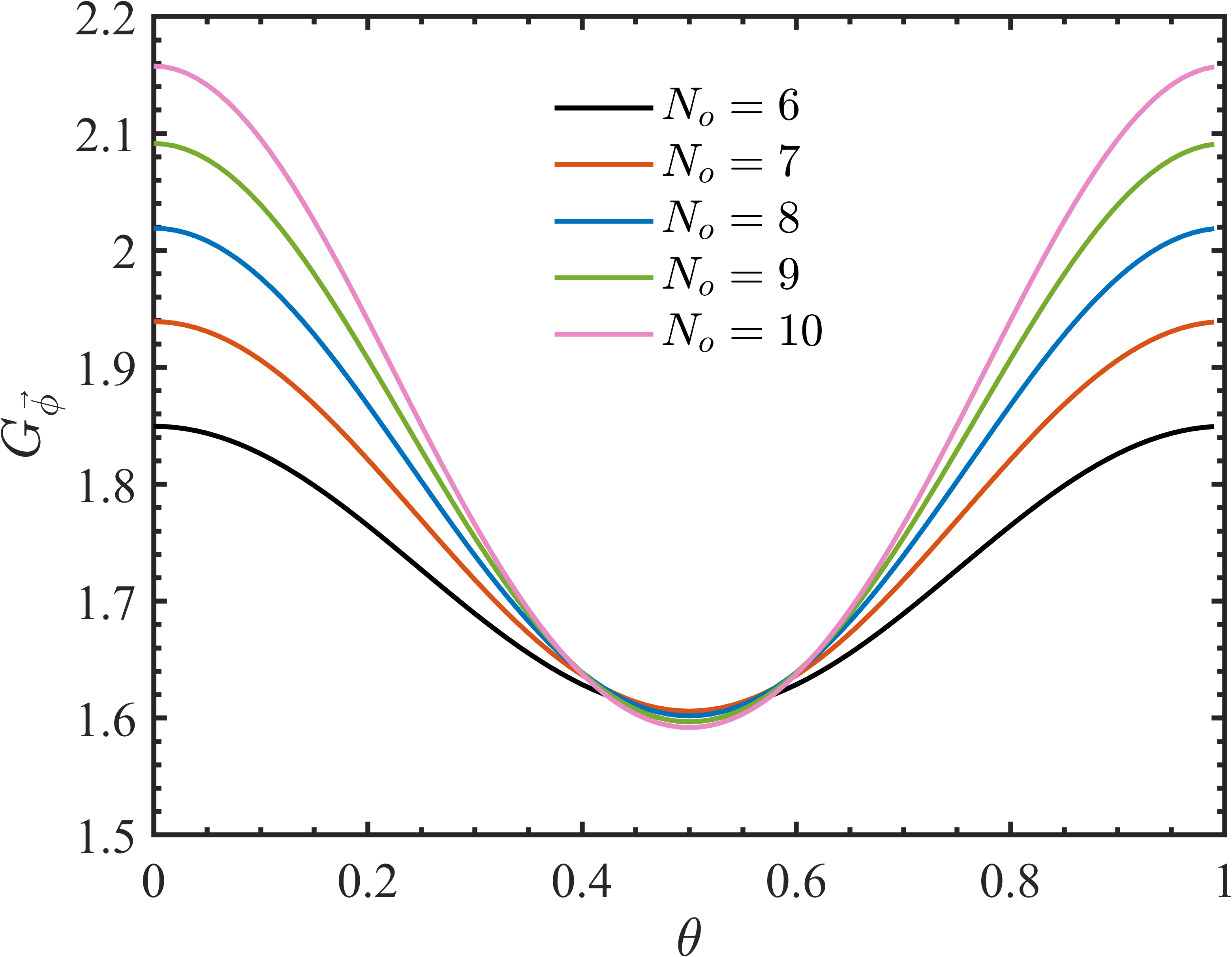}
\includegraphics[width=0.24\textwidth]{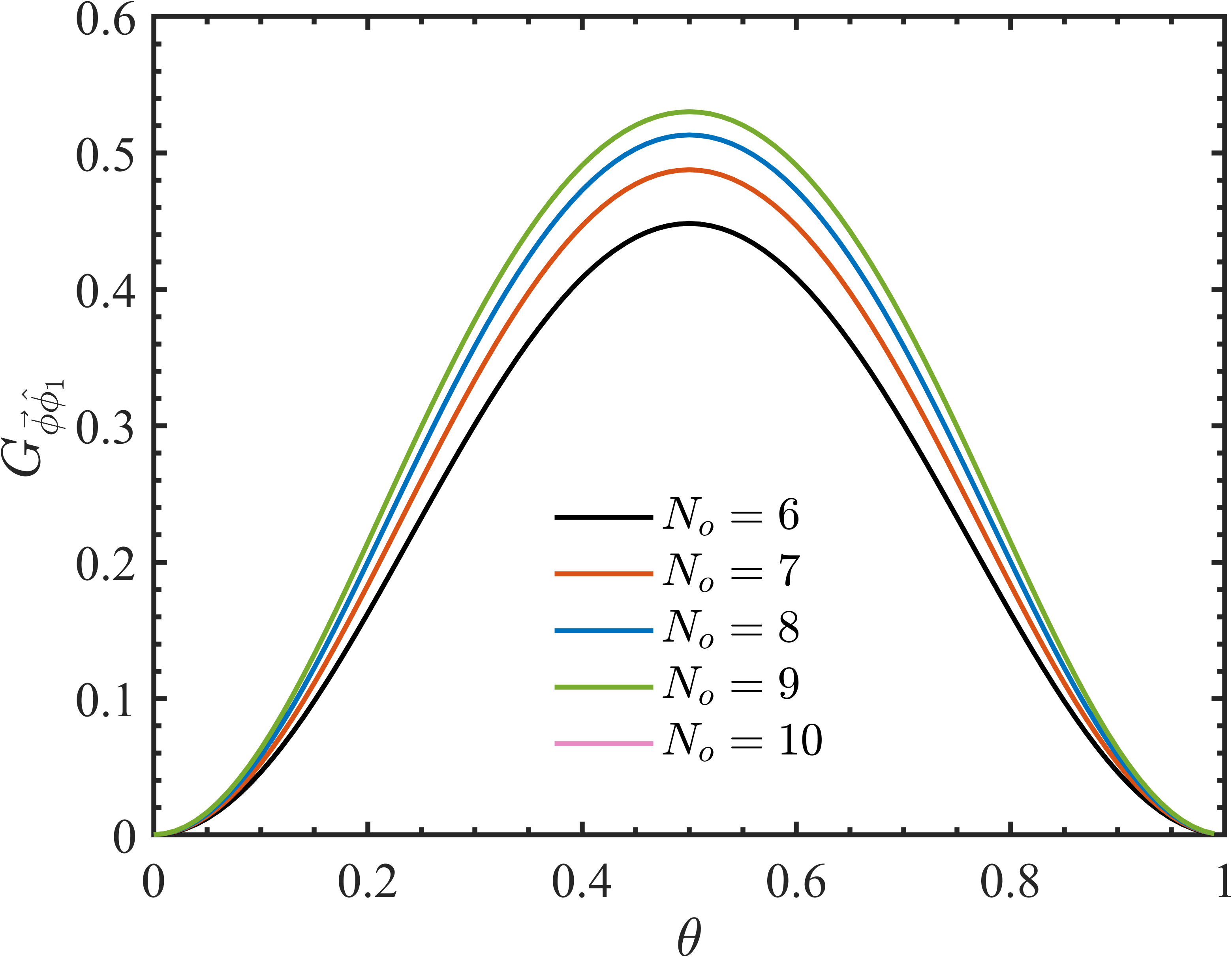}
\includegraphics[width=0.24\textwidth]{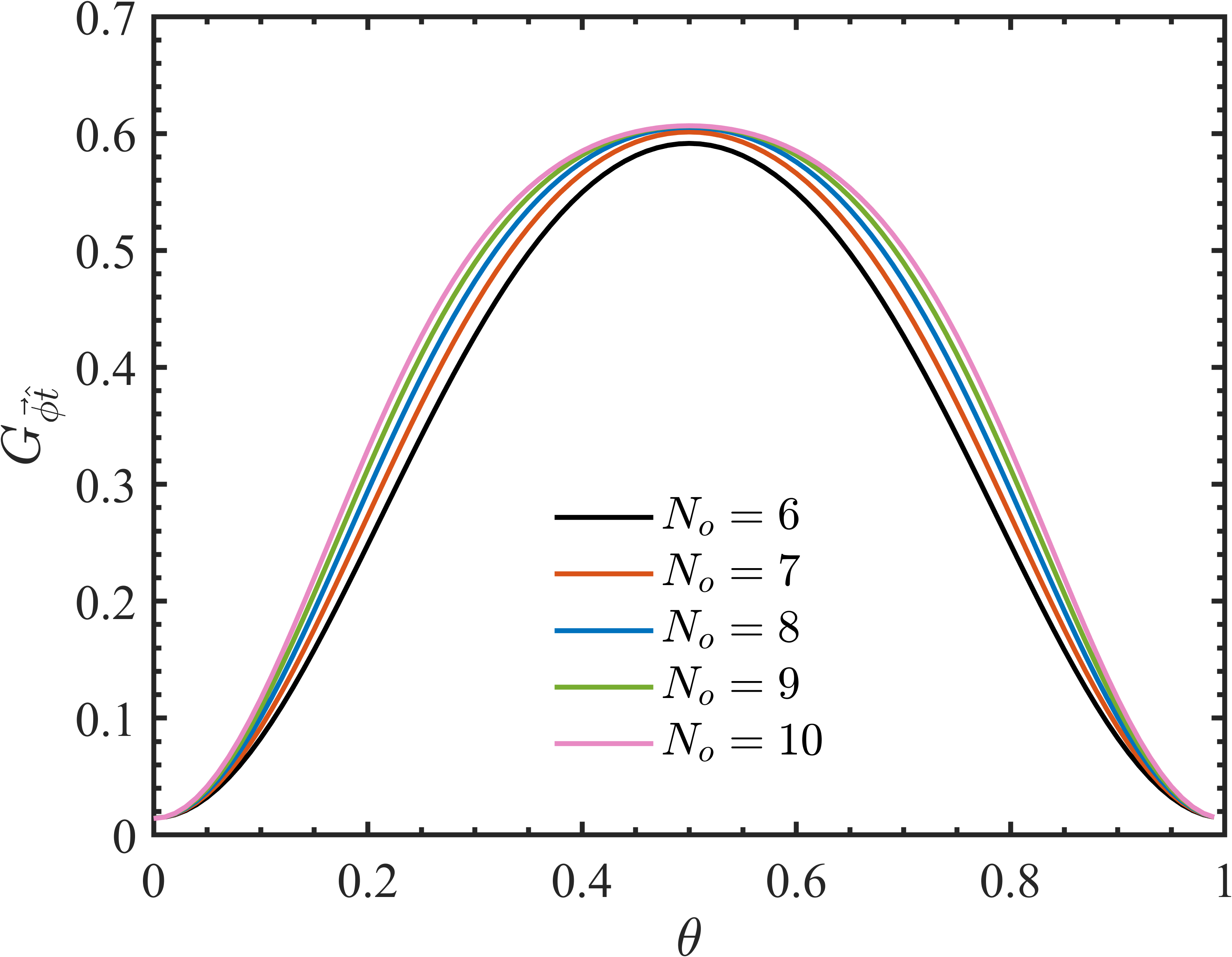}
\includegraphics[width=0.24\textwidth]{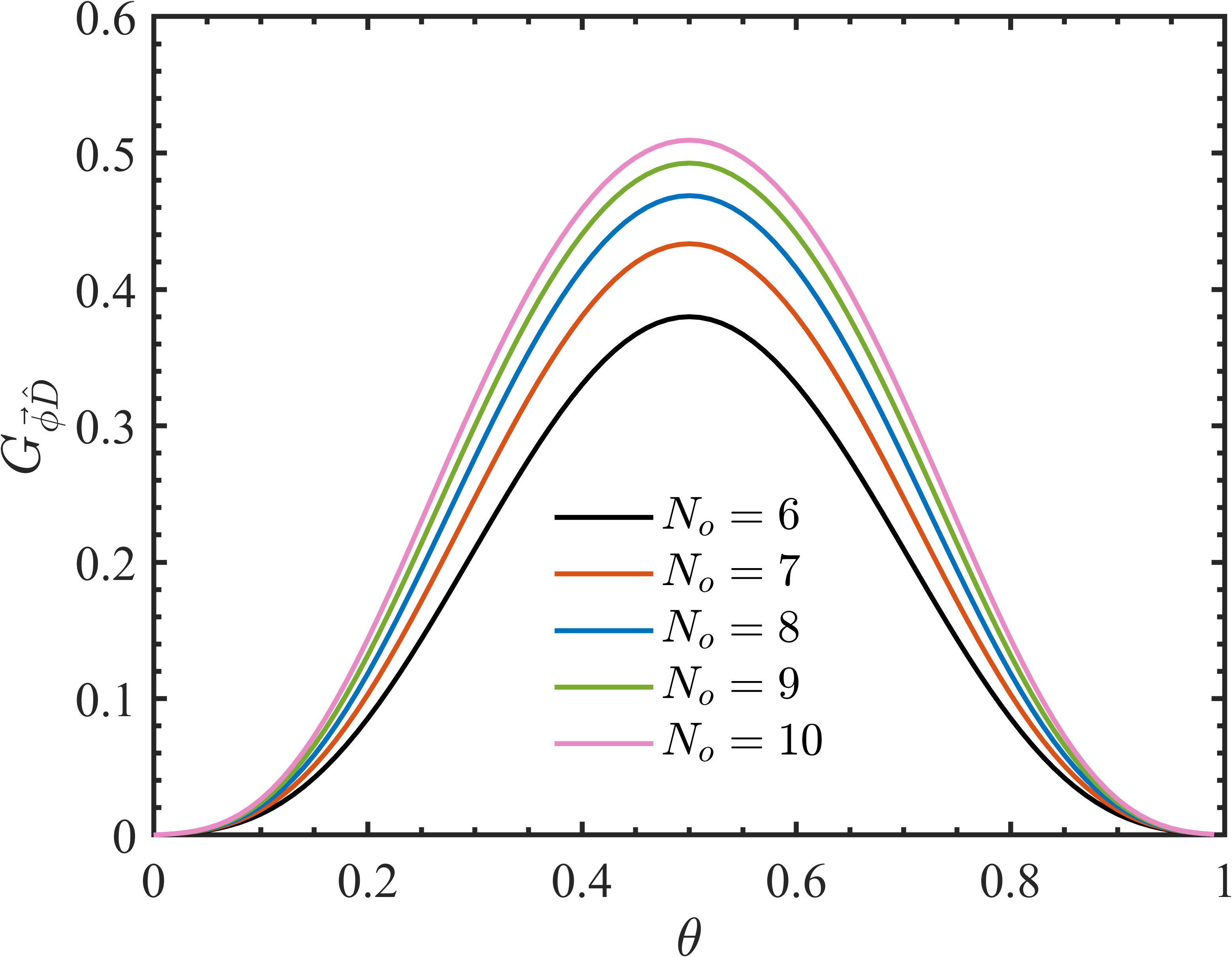}
    \caption{The one-point correlation function (a)$G_{\vec{\phi}}$ and two-point correlation function (b)$G_{\phi\hat{t}}$, (c)$G_{\phi\hat{D}}$ and (d) $G_{\phi\hat{t}}$ as a function of $\theta$. 
    As the system size increases, the numerical results gradually converge. These results are produced at $h_d=1000$ with subject to pseudo-critical SO($5$) bulk theories.
    }
    \label{fig:corr_func}
\end{figure*}

For this SO(5)-symmetric deconfined quantum critical point considered in this work, we numerically evaluate a set of such correlation functions on finite-size fuzzy spheres. For bulk fields $O(x)$, we mainly focus on the SO(5) vector operator $O=\vec{\phi}$. In numerics, we use the following density operator in the vector representation of SO(5) group to approximate $\vec{\phi}$:
\begin{equation}
    n_i(\boldsymbol{r})=\boldsymbol{\psi}^\dagger(\boldsymbol{r})\Gamma^i\boldsymbol{\psi}(\boldsymbol{r})
\end{equation}
Fig.~\ref{fig:corr_func} presents corresponding one-point and bulk-to-defect correlation functions as functions of the polar angle \(\theta\), which parametrizes the distance from the defect. The numerical data exhibit the scaling forms expected from defect conformal symmetry over a broad range of length scales. Moreover, around $\theta=\pi/2$, the results gradually approach the same values as the system size increases. 

\begin{figure*}[!htbp] 
    \centering
\includegraphics[width=0.24\textwidth]{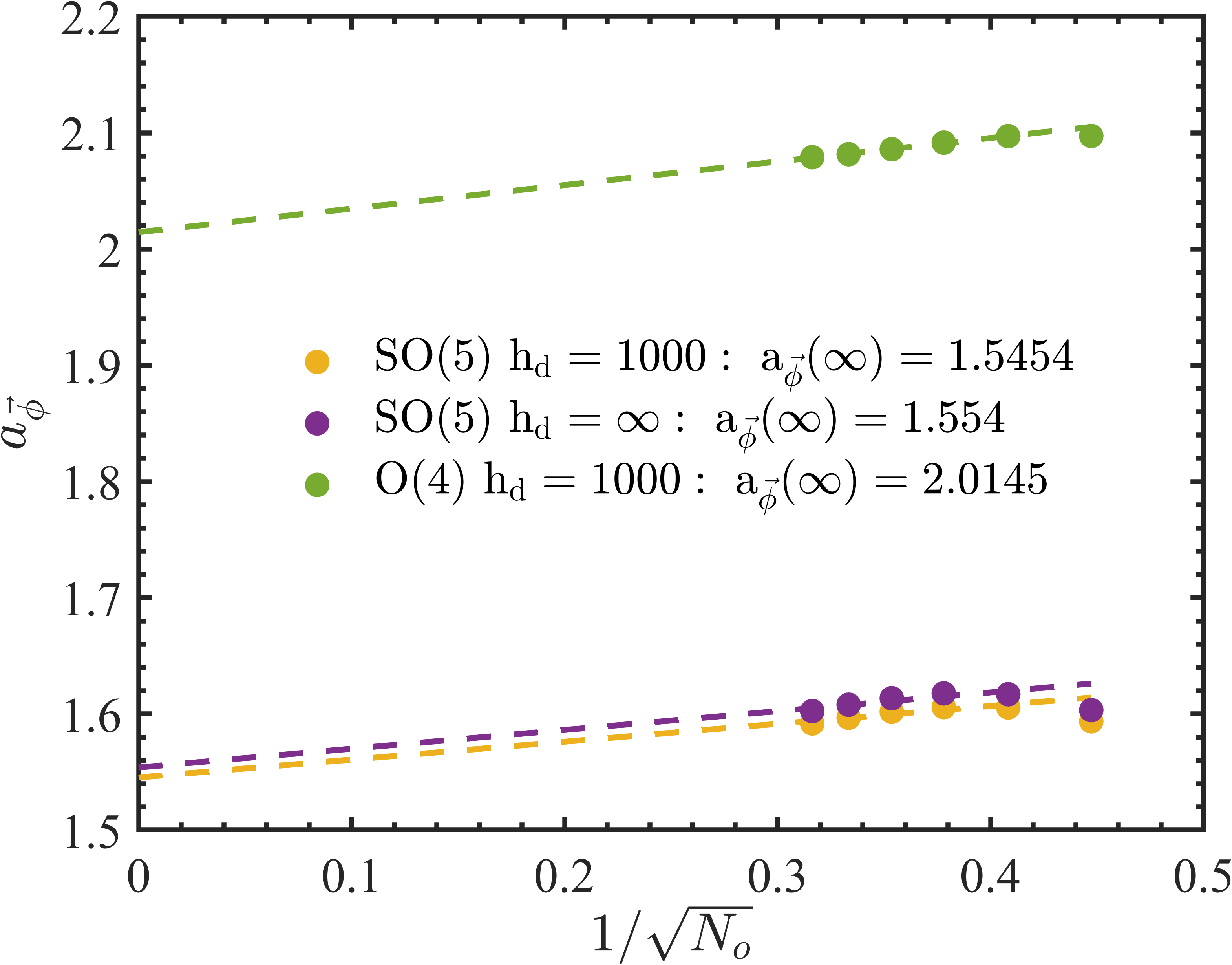}
\includegraphics[width=0.24\textwidth]{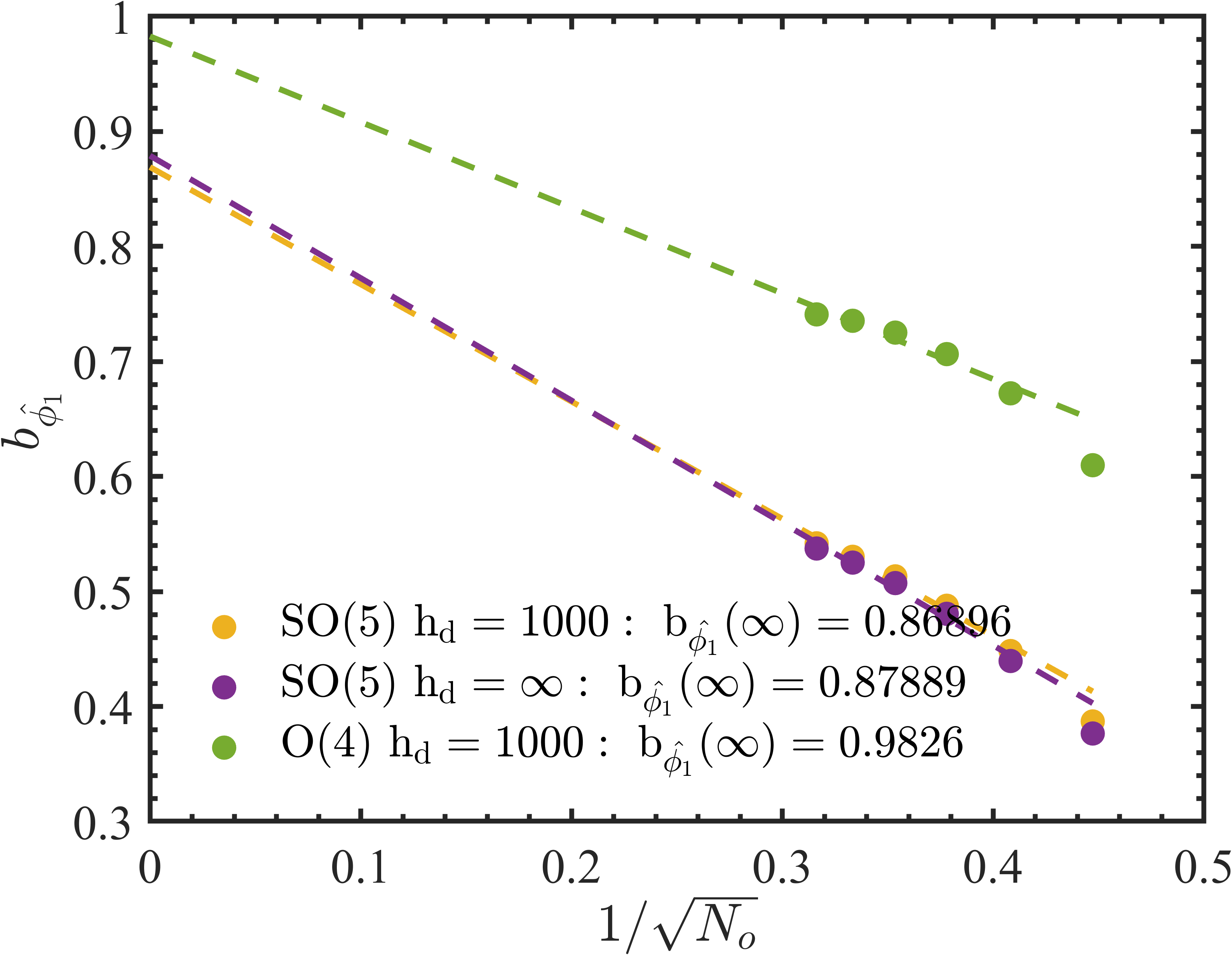}
\includegraphics[width=0.24\textwidth]{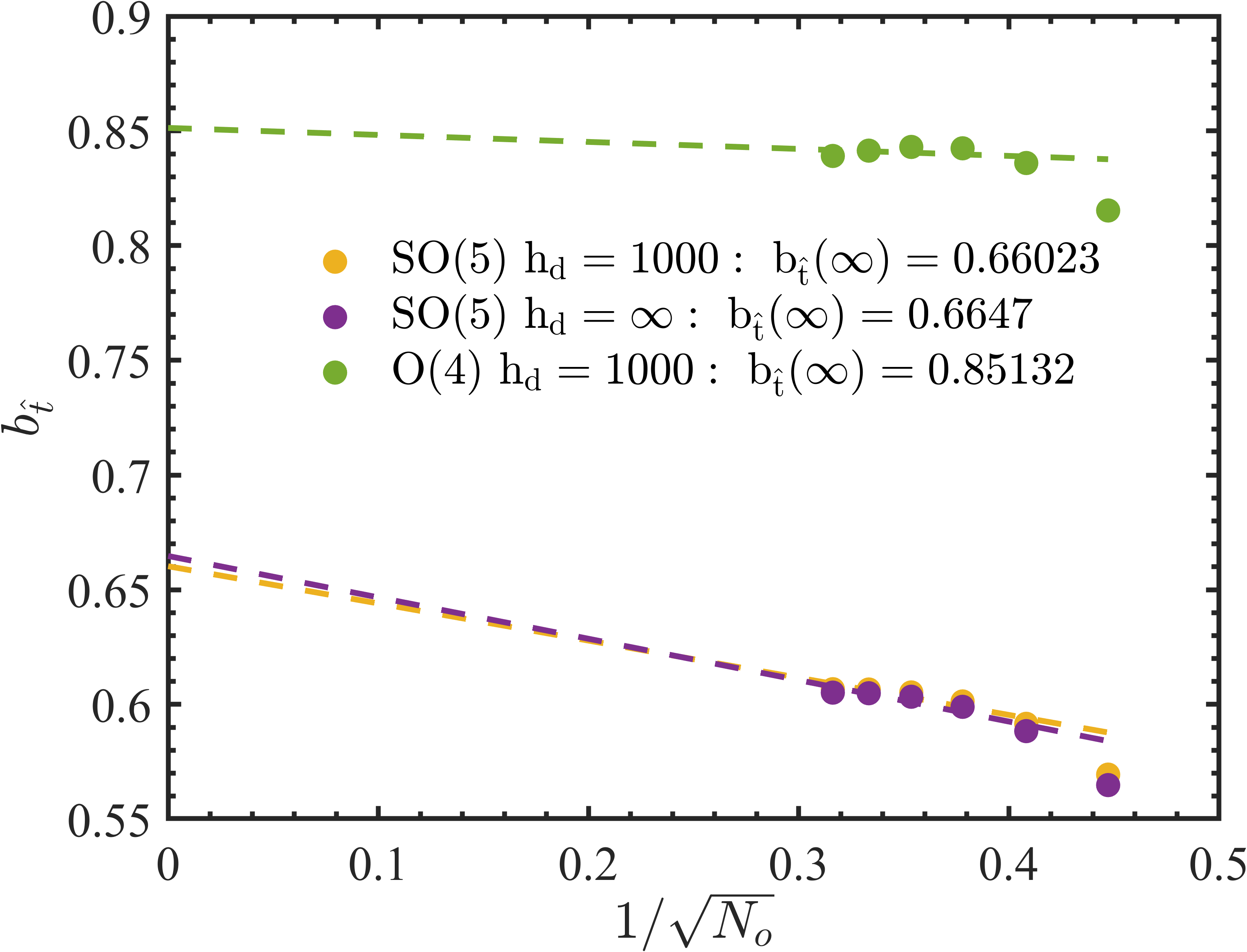}
\includegraphics[width=0.24\textwidth]{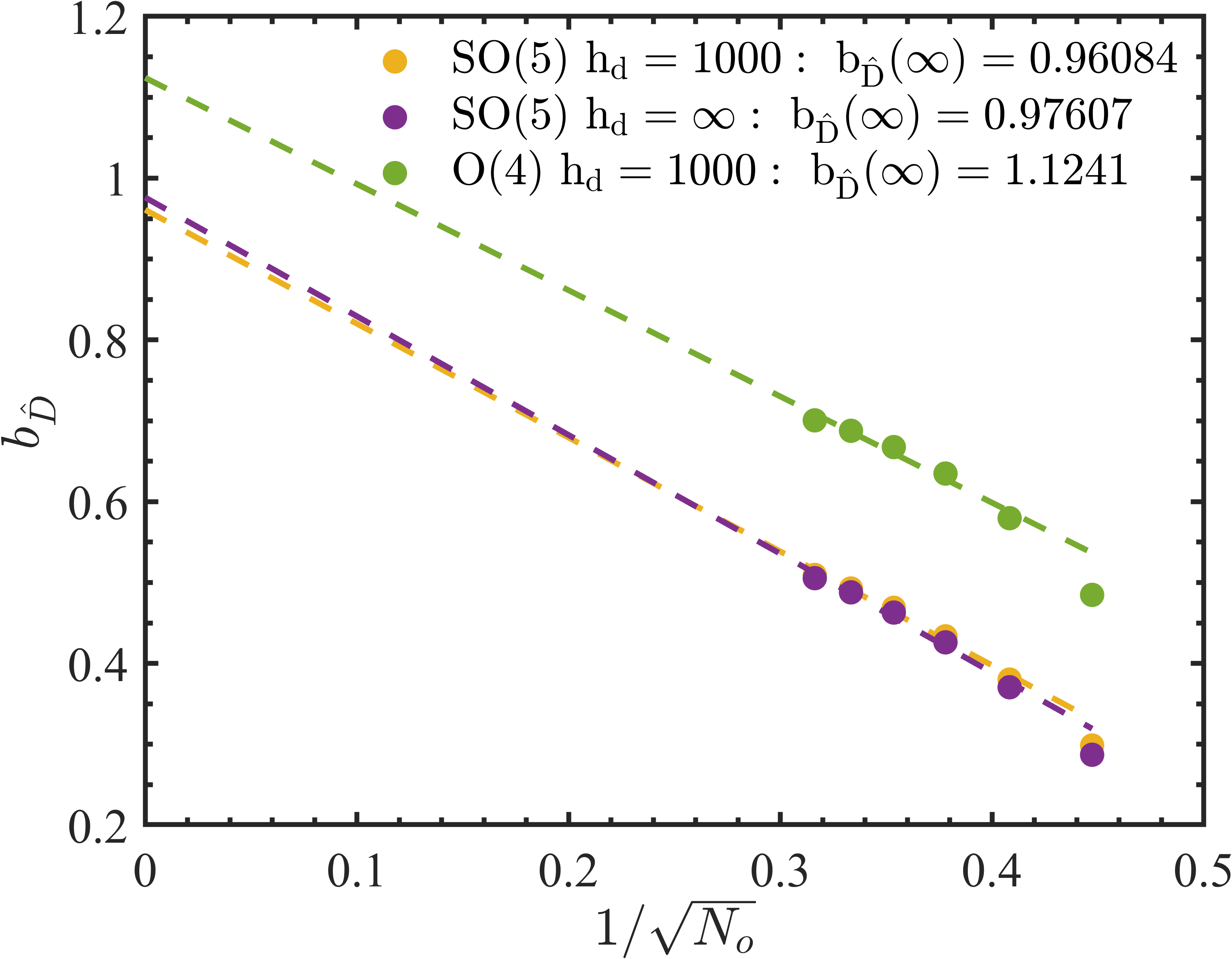}
    \caption{The finite-size extrapolation results of the one-point and two-point OPE coefficients
(a)$a_{\vec{\phi}}$, (b)$b_{\hat{\phi}_1}$, (c)$b_{\hat{t}}$ and (d)$b_{\hat{D}}$
calculated from bulk operator $\vec{\phi}$ in the
defect CFT fixed point by applying a spatially localized magnetic field $h_d$ subject to SO(5)/O(4) deconfined quantum critical bulk theory. Here, we adopt a linear fit to estimate these universal coefficients. For the same bulk fixed point, $h_d=1000$ and $h_d=\infty$ yield highly consistent results.
    }
    \label{fig:ope_extrapolate}
\end{figure*}

Using these finite-size data, we further perform extrapolations to the thermodynamic limit to obtain estimations of the universal coefficients \(a_{O}\) and \(b_{\hat{O}}\). The extrapolation procedures are illustrated in Fig.\ref{fig:ope_extrapolate}, and the final results are summarized in Table.\ref{tab:ope_and_scaling_dimension_result}.

\begin{table*}[!htbp]
\caption{At different bulk DQCPs, we summarize the finite-size extrapolation results of several universal OPE coefficients and primary operators' scaling dimensions by implying a large enough pinning field $h_d=1000$ or $h_d=\infty$.}
\begin{tabular}{c|c|cccc|ccccc}
\hline\hline
bulk symmetry & pinning field strength & $a_{\hat{\phi}}$ & $b_{\hat{\phi}}$ & $b_{\hat t}$ & $b_{\hat D}$ & $\Delta_{\hat{\phi}_a}$ & $\Delta_{\hat{\phi}_1}$ & $\Delta_{\partial_\mu\hat{\phi}_1}$ & $\Delta_{\hat{\phi}^{+-}}$ & $\Delta_{\hat{\phi}^{+0}}$ \\
\hline
\multirow{2}{*}{SO(5)} 
& $h_d=1000$   & 1.5454 & 0.8690 & 0.6602 & 0.9608 & 0.9764 & 2.0382 & 2.7861 & 1.3939 & 0.2413 \\
& $h_d=\infty$ & 1.5540 & 0.8789 & 0.6647 & 0.9761 & 0.9665 & 1.9822 & 2.8651 & 1.4222 & 0.2207 \\
\hline
O(4) 
& $h_d=1000$   & 2.0145 & 0.9826 & 0.8513 & 1.1241 & 0.9706 & 1.2614 & 2.5729 & 1.8603 & 0.1819 \\
\hline\hline
\end{tabular}
\label{tab:ope_and_scaling_dimension_result}       
\end{table*}

\subsection{ Spectrum of defect creation and changing operators}
Besides the operators discussed above, defect CFT also contains another class of local operators associated with changes of the defect type. If we insert different defect types \(\mathcal{D}^a\) and \(\mathcal{D}^b\) at the north and south poles, respectively, we denote the corresponding Hamiltonian by the defect-changing Hamiltonian \(H^{ab}\). Under the Weyl transformation implementing the state–operator correspondence between \(\mathbb{R}\times S^{D-1}\) and \(\mathbb{R}^D\), this defect configuration is mapped to a straight infinite line defect in \(\mathbb{R}^D\) containing a localized junction at the origin, across which the defect type changes. The operators localized at such junctions are known as defect-changing operators \(\hat{\phi}^{ab}\). This construction is analogous to the boundary-condition-changing operators in two-dimensional boundary conformal field theories\cite{CARDY1989581}, which interpolate between different boundary conditions. 

\begin{figure}[!htbp] 
    \centering
\includegraphics[width=0.2\textwidth]{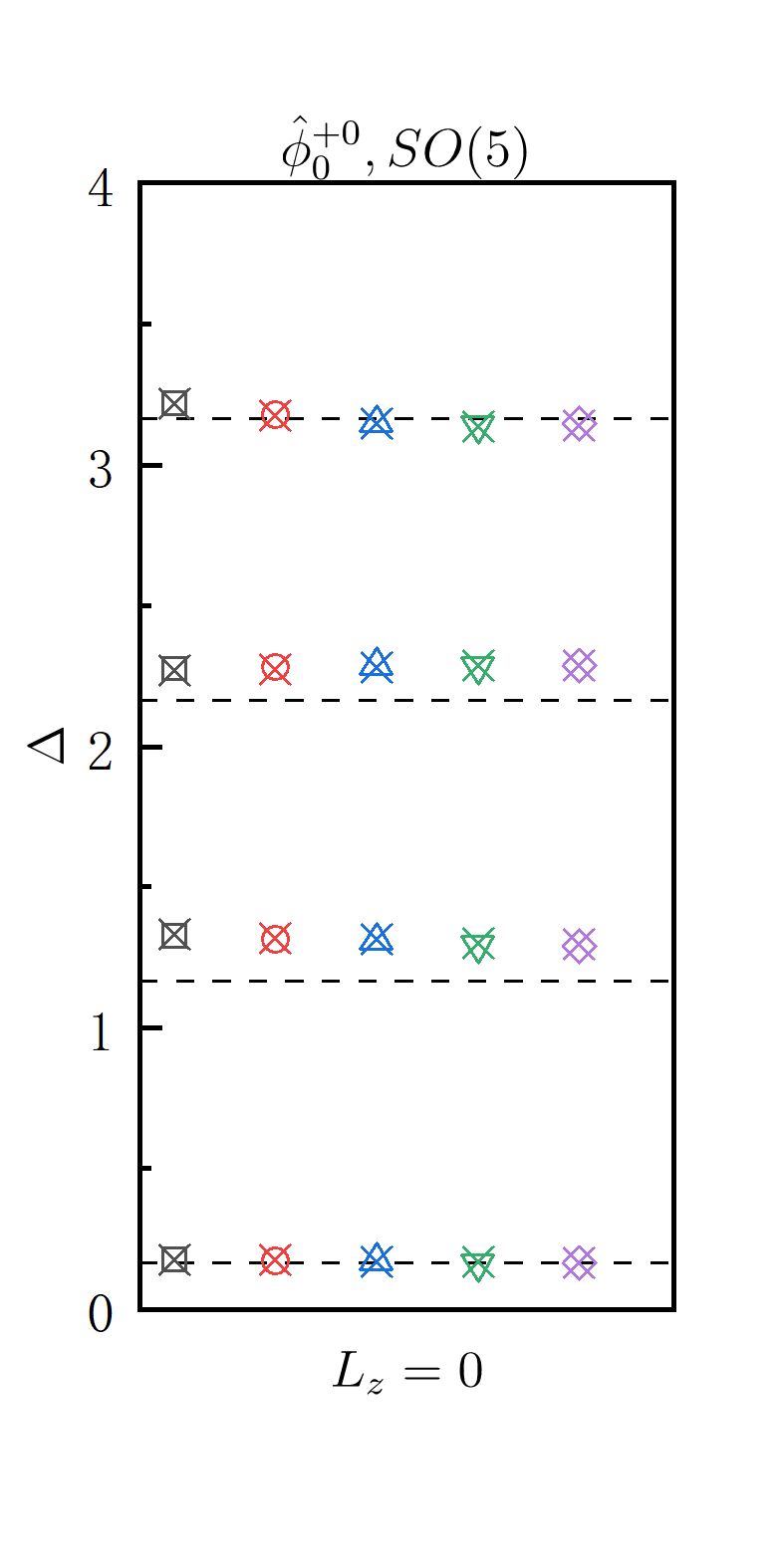} 
\includegraphics[width=0.2\textwidth]{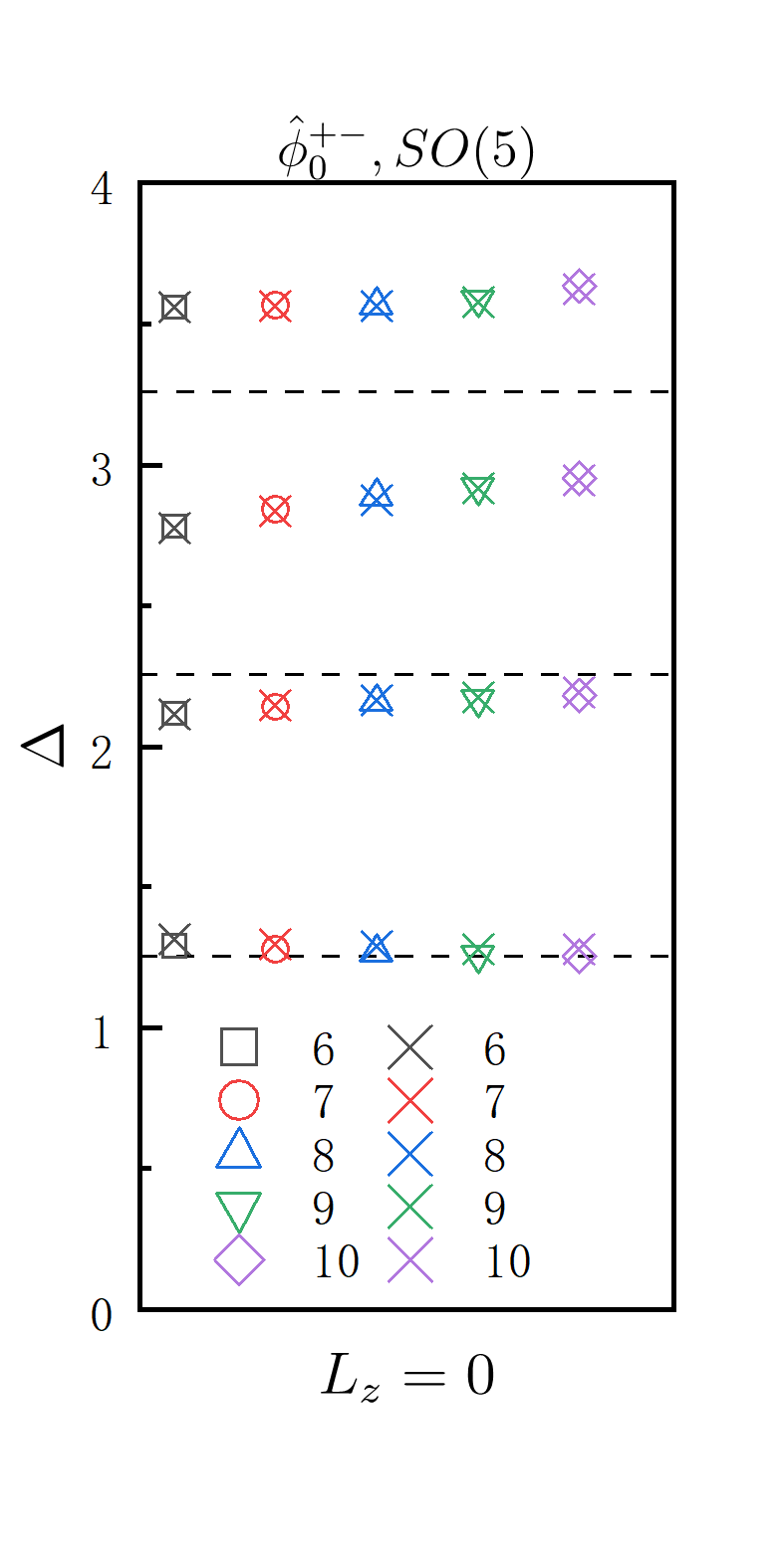}
    \caption{Multiplet structures of the (a) defect-creating and (b) defect-changing operators, corresponding to the ground states of the defect-creating and defect-changing Hamiltonians, respectively. The results are shown for pinning field strengths $h_d=1000$ and $h_d=\infty$(cross symbols), respectively, when the bulk is tuned to be at SO(5) deconfined quantum critical point\cite{Zhou_2024}.
    }
    \label{fig:so5_real_pm_p0_multi}
\end{figure}

A closely related object is the defect-creating operator $\hat{\phi}^{a0}$, which interpolates between the trivial bulk (identity defect) and a nontrivial defect $\mathcal{D}^{a}$. Both defect-changing and defect-creation operators are local operators living on the defect worldline and organize into conformal multiplets under the preserved one-dimensional conformal group $SL(2,\mathbb{R})$. The primary operators are determined by the energy gaps of leading excited states in the corresponding defect Hamiltonian. We show the corresponding conformal towers extracted from the numerical spectrum in Fig.~\ref{fig:so5_real_pm_p0_multi}.
Furthermore, we perform a self-consistent finite-size scaling analysis to extrapolate the scaling dimensions of the defect-changing operators and defect-creating operators to the thermodynamic limit.

\begin{figure}[!htbp] 
    \centering
    \includegraphics[width=0.8\linewidth]{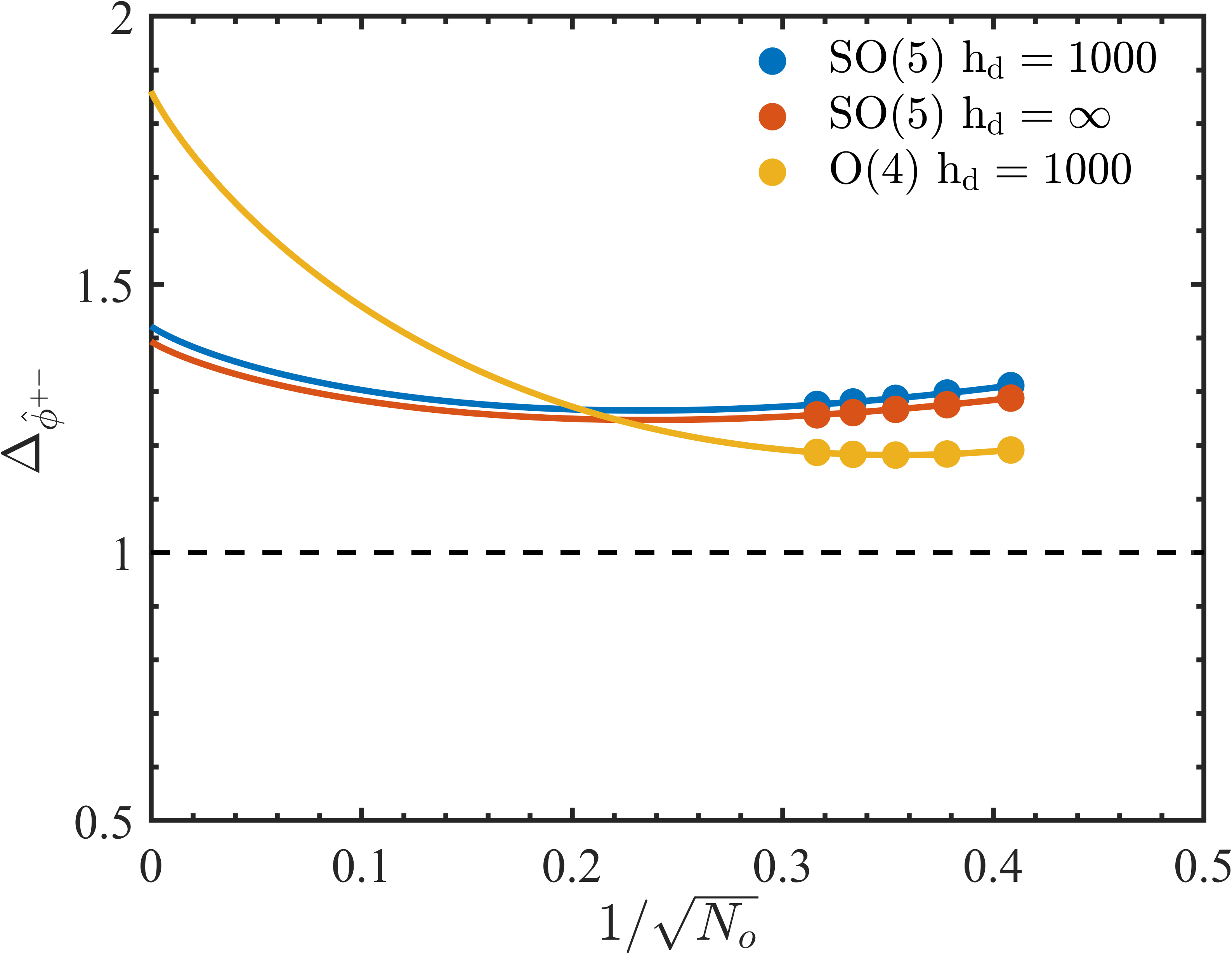}
    \includegraphics[width=0.8\linewidth]{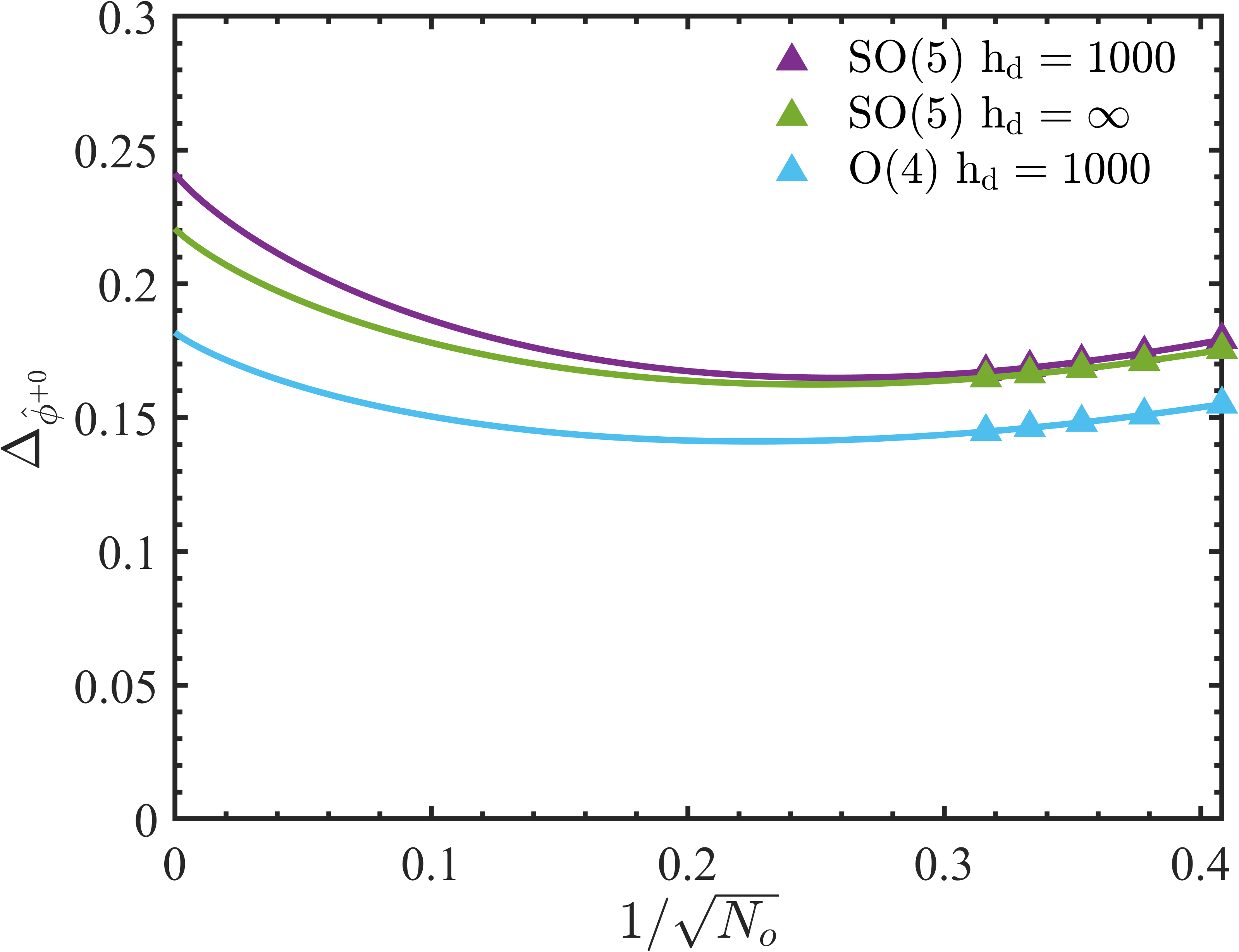}
    \caption{The finite-size extrapolation of defect (a) changing operator $\hat\phi^{+-}$ and (b) creating operator $\hat\phi^{+0}$ at different deconfined quantum critical bulk. We extrapolate to the thermodynamic limit through the scaling form of Eq.\ref{eq:finite_size_consistent_scaling}. 
    }
    \label{fig:phi_pm_and_phi_p0_extrapolate_diff_ON_dqcp}
\end{figure}

In our analysis, we primarily account for corrections arising from the two lowest-lying irrelevant scalar operators. I.e. we use the following scaling form:
\begin{equation}
    \Delta_{\hat{O}}(R)=
\Delta_{\hat{O}}(\infty)
+
\frac{b}{R^{\Delta_{\hat{\phi}_1}-1}}
+
\frac{c}{R^{\Delta_{\hat{S}^-}-1}}.
\label{eq:finite_size_consistent_scaling}
\end{equation}
By imposing self-consistency between the extrapolated scaling dimensions and the fitted correction exponents, we obtain reliable estimates of the thermodynamic-limit values. The resulting extrapolations are summarized below in Tab.\ref{tab:ope_and_scaling_dimension_result}, while the detailed fitting procedures and numerical data are shown in Fig.\ref{fig:phi_pm_and_phi_p0_extrapolate_diff_ON_dqcp}.

Here, the defect-changing operator admits a simple physical interpretation under the assumption that the infrared defect fixed point is described by a non-simple defect $\mathcal{D}^+\oplus \mathcal{D}^-$, corresponding to a possible phase with spontaneously broken $Z_2$ symmetry. In this picture, the two elementary defects $\mathcal{D}^+$ and $\mathcal{D}^-$ represent the two symmetry-related ordered sectors, while the defect-changing operator is naturally identified with the domain-wall operator connecting them. Consequently, its scaling dimension determines the stability of the ordered defect: if $\Delta_{\hat{\phi}^{+-}}>1$,
the domain-wall perturbation is irrelevant, implying that the defect long-range order remains stable in the infrared. The finite-size extrapolation shown in Fig.\ref{fig:phi_pm_and_phi_p0_extrapolate_diff_ON_dqcp}(a) yields the following estimate for the scaling dimension of the domain-wall operator on the line defect of the DQCP:
\begin{equation}
    \Delta_{\hat\phi^{+-}}^{SO(5)/O(4)}(\infty)>1
\end{equation}
 This result therefore supports the existence of a stable symmetry-breaking phase on the one-dimensional line defect embedded in the deconfined quantum critical bulk.  Whether a one-dimensional defect can sustain spontaneous symmetry breaking has recently attracted considerable attention. Recent studies of line defects in the three-dimensional Ising CFT based on the fuzzy sphere approach\cite{dCFT_Sphere_Zhou_2024}, together with conformal bootstrap calculations\cite{lanzetta2025beginningendpointbootstrapconformal}, have shown that in the Wilson-Fisher CFT with a global $Z_2\sim O(1)$ symmetry, the coupling to the critical bulk cannot induce sufficiently long-ranged interactions along a one-dimensional line defect to stabilize spontaneous symmetry breaking. In contrast, for line defects in the O(N) Wilson-Fisher CFT, it has been proposed that there exists a critical value $N_c$, such that stable long-range order can develop on the one-dimensional defect when $N>N_c$\cite{Cuomo_2022_epsilon_on_defects}. For line defects embedded in a deconfined quantum critical point, however, this issue has remained unexplored. The results reported in this work provide the first systematic investigation of this problem. In the following, we will investigate this issue through the calculation of the defect $g$-function.
 
\subsection{Defect $g$-function}
\begin{figure}[!htbp] 
    \centering
    \includegraphics[width=0.8\linewidth]{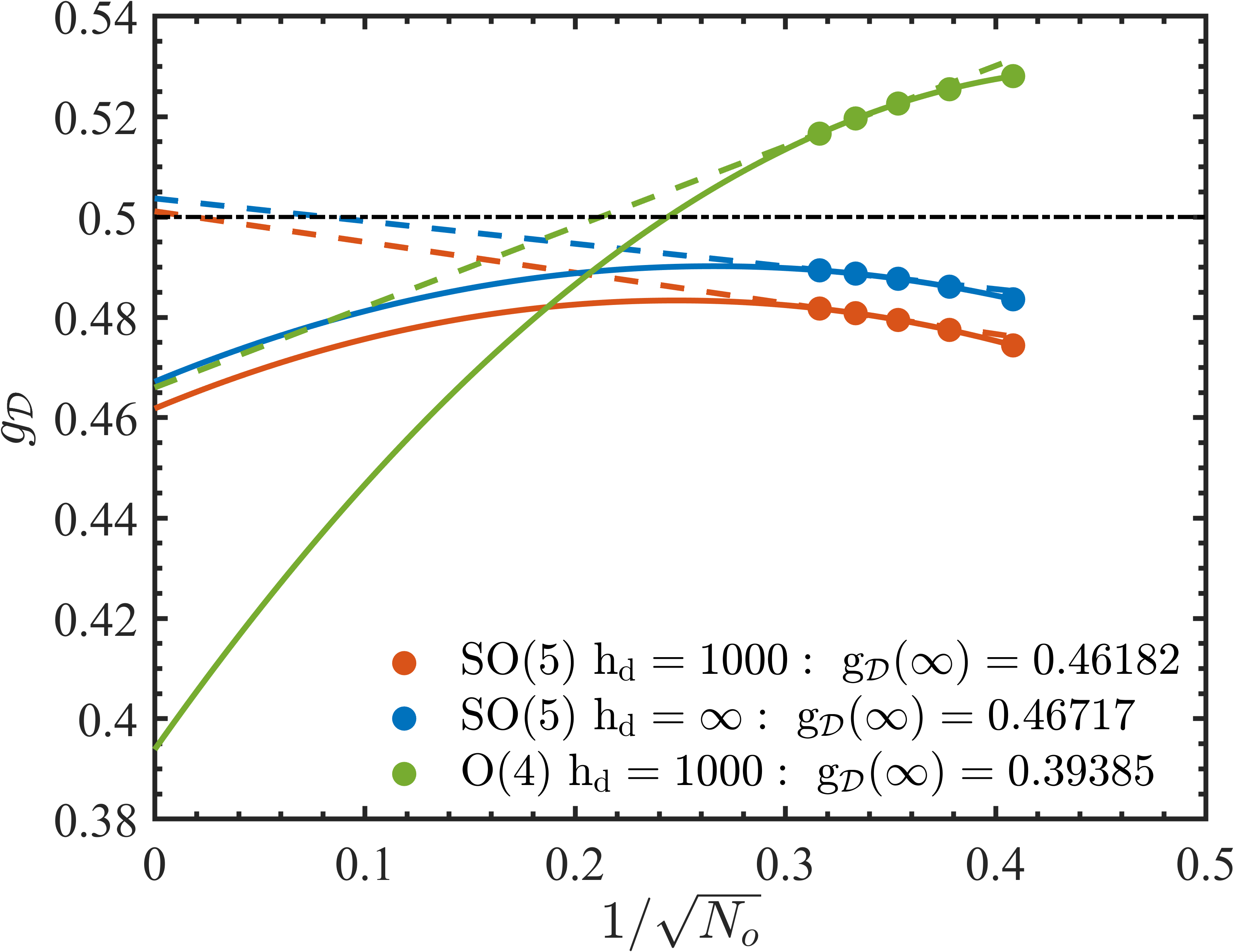}
    \caption{The finite-size extrapolation of the defect $g-$function at different bulk deconfined quantum critical points. We extrapolate to the thermodynamic limit through scaling form of $g_{\mathcal{D}}(N_o)=g_{\mathcal{D}}(\infty)+aN_o^{-1/2}+bN_o^{-1}$. We also present the results obtained from linear fits (dashed lines). The polynomial fitting schemes lead to the conclusions $g_{\mathcal{D}}<0.5$, while the linear fitting gives a value slightly larger than 0.5. We consider the polynomial fit to be more reliable, since the finite-size data exhibit a nonlinear behavior. 
    }
    \label{fig:phi_pm_g_func_extrapolate_diff_ON_dqcp}
\end{figure}

Another important piece of conformal data associated with a defect fixed point is the defect $g$-function\cite{Affleck_Ludwig_g_theorem_PhysRevLett.67.161}. According to the $g$-theorem, the value of $g$-function decreases monotonically along RG flows connecting defect fixed points\cite{Affleck_Ludwig_g_theorem_PhysRevLett.67.161,Casini_2016_gTheorem}. 
Equivalent definitions of the g-function exist from both the quantum field theory and quantum information perspectives\cite{Casini_2016_gTheorem,Cuomo_2022}. 
Within the fuzzy sphere formulation, this universal quantity can be measured through a practical procedure in numerical simulations. Specifically, the evaluation of the $g$-function only involves the overlaps between the ground-state wave functions $|ab\rangle$($a,b$ represent different defects inserted on the north pole and south pole) of Hamiltonians with different defect configurations:
\begin{equation}
        g=\frac{|\langle+0|00\rangle|^2}{|\langle+0|++\rangle|^2}
    \label{eq:g_func_ovlap}
\end{equation}
For the detailed derivation of Eq.\ref{eq:g_func_ovlap}, we refer the reader to the literature\cite{dCFT_Sphere_Zhou_2024}. One can immediately recover that $g_{\text{bulk}}=1$ in the absence of any defect insertion, which corresponds to the trivial identity defect. For the defect fixed points considered in this work, the computed values of the g-function are shown in the Fig.\ref{fig:phi_pm_g_func_extrapolate_diff_ON_dqcp}.
To estimate its thermodynamic-limit value, we perform polynomial extrapolations as a function of system size. As shown in Fig.\ref{fig:phi_pm_g_func_extrapolate_diff_ON_dqcp}, the line defects at all deconfined quantum critical points considered in this work satisfy
\[
g^{SO(5)/O(4)}_{\mathcal{D}}(\infty)<0.5<g_{\mathrm{bulk}}=1.
\]
This result provides further evidence for the existence of a non-trivial defect fixed point rather than being screened by the bulk. More importantly, $g_{\mathcal{D}}<0.5$ provides useful information regarding the possibility of spontaneous symmetry breaking on the defect. Suppose that the infrared fixed point corresponds to a symmetry-breaking phase described by the non-simple defect $\mathcal D^+\oplus\mathcal D^-$, 
where $\mathcal D^\pm$ represent the two symmetry-related ordered sectors. Since the $g$-function is additive under direct sums of defects,
$g_{\mathcal D^+\oplus\mathcal D^-}
=2g_{\mathcal D}$, 
the monotonicity implied by the $g$-theorem requires $2g_{\mathcal D}<g_{\rm bulk}=1$,
which immediately gives $g_{\mathcal D}<0.5$\footnote{Strictly speaking, the condition $g_{\mathcal{D}}<0.5$ is only meaningful for assessing the stability of the spontaneous-symmetry-breaking fixed point associated with this particular symmetry-breaking pattern. For other possible spontaneous symmetry-breaking scenarios, this condition is neither necessary nor sufficient.}.
Therefore, the observed inequality $g_{\mathcal{D}}<0.5$
is fully consistent with the scenario that the infrared defect fixed point is a symmetry-breaking defect.

Combined with the scaling analysis of the defect-changing operator presented in the previous subsection, the $g$-function further sheds light on the stability of symmetry breaking on the defect. Our numerical evidence demonstrates that a one-dimensional line defect coupled to a DQCP can indeed develop stable long-range order through spontaneous symmetry breaking\cite{Mermin_Wagner_PhysRevLett.17.1133}. This is particularly noteworthy because an isolated one-dimensional system cannot exhibit spontaneous symmetry breaking\cite{Dyson:1968up,Hohenberg_PhysRev.158.383,Anderson_PhysRevB.1.4464,Kosterlitz_PhysRevLett.37.1577}, whereas a defect embedded in a higher-dimensional critical bulk can evade this restriction through its coupling to the gapless bulk degrees of freedom\cite{Max_boundary_10.21468/SciPostPhys.12.4.131,Krishnan_2023,lanzetta2025beginningendpointbootstrapconformal}.


\section{Summary and discussion}
\label{sec:summ_and_diss}
In this work, we have presented the first systematic study of one-dimensional line defects embedded in two-dimensional deconfined quantum critical points within the framework of defect conformal field theory. Using the fuzzy-sphere regularization, we characterized the infrared defect fixed point through a broad set of conformal data, including the defect operator spectrum, bulk-to-defect correlation functions, OPE coefficients, defect-creation and defect-changing operators, and the defect \(g\)-function. The emergence of well-defined defect conformal multiplets provides strong evidence that the pinning-field perturbation flows to a nontrivial defect conformal fixed point.

Our results further reveal an intriguing interplay between topology and defect criticality. We found that a DQCP bulk appears to generically admit a stable symmetry-breaking on the defect in contrast to conventional O(N) Wilson–Fisher bulk CFTs\cite{dCFT_Sphere_Zhou_2024,lanzetta2025beginningendpointbootstrapconformal}. The finite-size extrapolations indicate that the pinning-field defect is not screened by the critical bulk, but instead flows to a stable infrared defect fixed point with its own conformal spectrum. Moreover, the irrelevance of the defect-changing (domain-wall) operator together with the value of the defect \(g\)-function consistently supports the scenario that the infrared defect supports stable long-range order along the one-dimensional defect. This result is particularly remarkable because an isolated one-dimensional system cannot spontaneously develop long-range order, whereas coupling to a higher-dimensional gapless critical bulk can effectively generate sufficiently long-ranged interactions to evade this restriction. Finally, we emphasize that a more definitive resolution of this issue would require identifying a potential SSB fixed point directly. This may be achieved by introducing an appropriate $Z_2$-symmetric line-defect perturbation, whose RG flow could reveal the existence of such a fixed point.

An important limitation of the present work arises from finite-size effects. First, the putative deconfined quantum critical points in the bulk theories considered here are unlikely to correspond to genuine continuous phase transitions, but are instead more likely weakly first-order transitions exhibiting pronounced pseudocritical behavior. Second, the determination of the bulk critical point can influence the extraction of conformal data at the defect fixed point. In particular, the location of the candidate O(4)-symmetric deconfined quantum critical point studied here is affected by several gapped excitations in the energy spectrum\cite{ys_o4dqcp_l6vw-6z79}. Finally, the extrapolation of conformal data extracted from finite-size systems to the thermodynamic limit inevitably introduces additional uncertainties. Together, these factors constitute the primary sources of the finite-size effects in the conformal data reported in this work. Developing systematic approaches to control and eliminate these effects, thereby enabling a more precise determination of defect conformal data, remains an important direction for future research.

Several interesting directions deserve further investigation. First, one may study codimension-one defects, namely surface critical phenomena at the bulk DQCPs\cite{Ma_2022,JYLee_PRXQuantum.4.030317}. Recent Monte Carlo simulations have already begun to explore such questions and revealed a variety of unconventional boundary universality classes\cite{zhu_yanzhang_tkx5-kzhh}. Second, it would be interesting to consider more general quantum impurities, such as spin-\(s\) impurities\cite{komargodski2025defectanomaliesspinfluxduality} coupled to a DQCP bulk, and determine their infrared fixed points and screening behavior. Finally, one may investigate defect universality classes enriched by additional gapless boundary degrees of freedom, such as edge modes inherited from symmetry-protected topological phases or gapless fermionic excitations\cite{toldin2025extraordinarytransitionedgecorrelated,edge_dqcp_liu2025edgemodestopologicalmott,tjfk-84f8,4lv4-mc81}.  Extending the defect-CFT program developed here to these systems may provide a fruitful route toward understanding the defect degrees of freedom in higher-dimensional gapless bulk due to the interplay among topology and criticality.



\begin{acknowledgements}
   We are grateful to Yin-Chen He, Shang Liu, Liang-dong Hu and Wei Zhu for the fruitful discussion. This work are supported by the National Key Research and Development Program of China Grant No. 2022YFA1402204 and the National Natural Science Foundation of China Grant No. 12274086.
\end{acknowledgements}

 \appendix
\section{line defects in O(4) deconfined quantum phase transition}\label{appsec:defect_in_o4_dqcp}
\begin{figure*}[!htbp] 
    \centering
\includegraphics[width=0.2\textwidth]{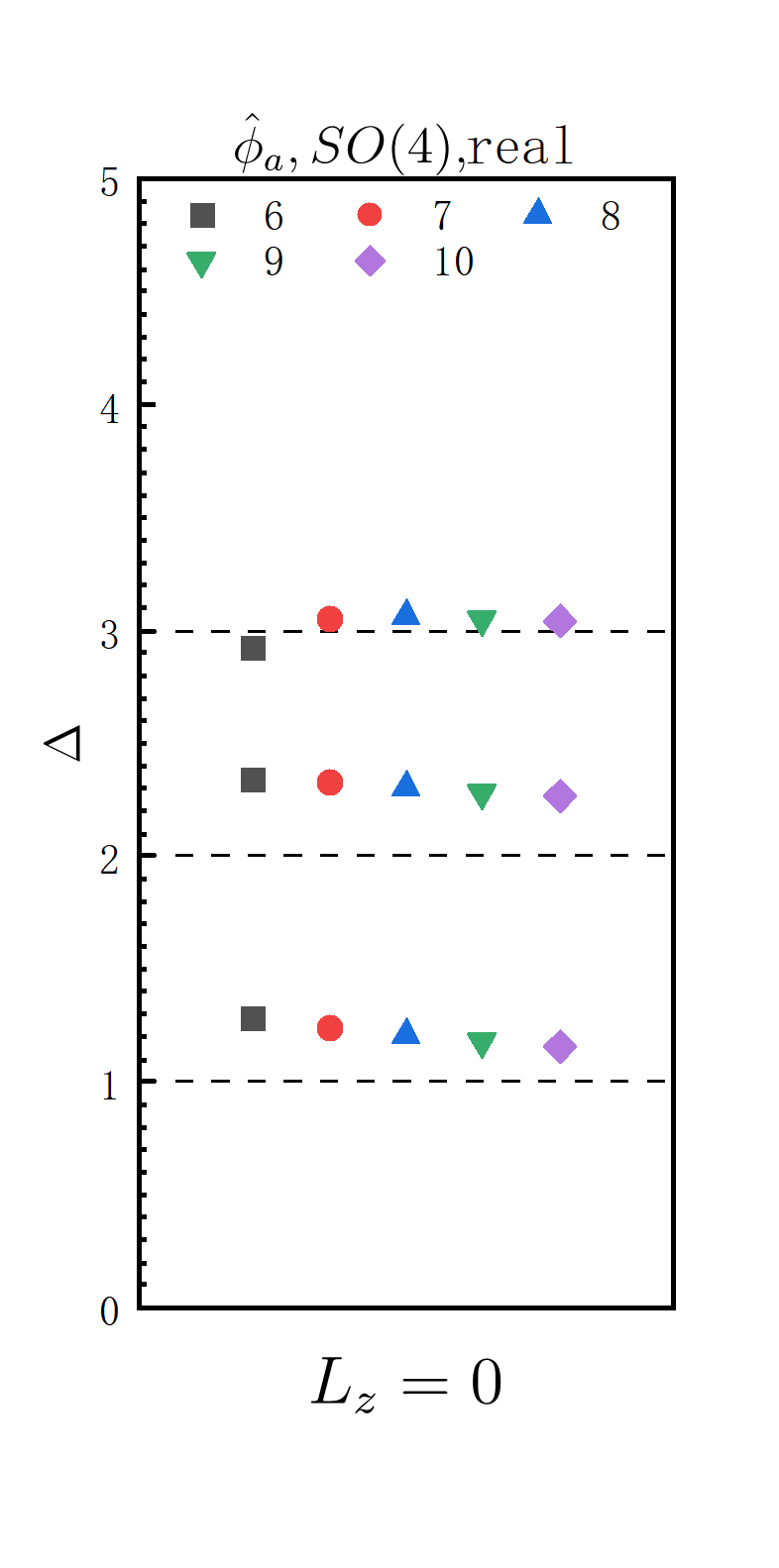} 
\includegraphics[width=0.2\textwidth]{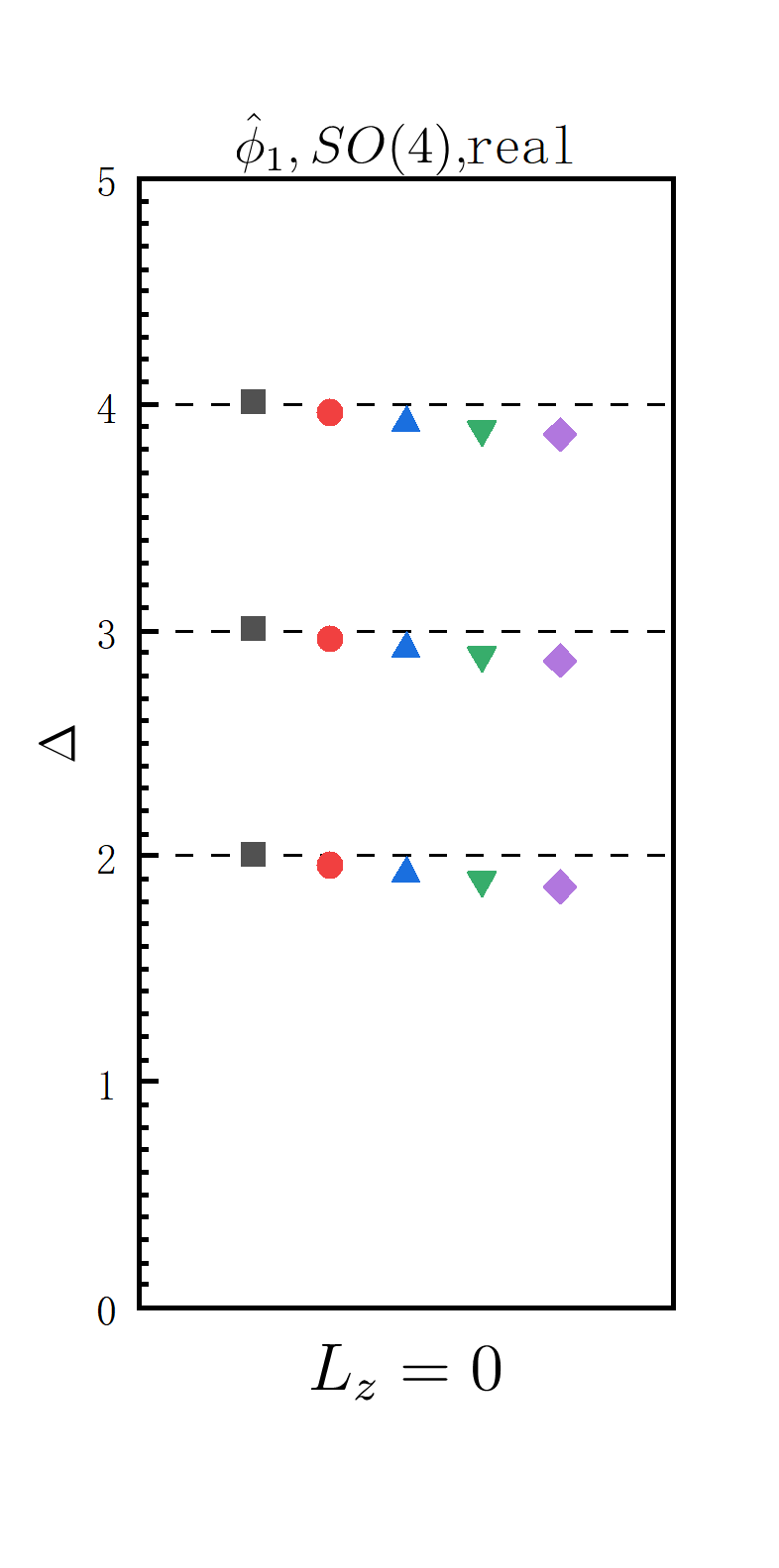}
\includegraphics[width=0.2\textwidth]{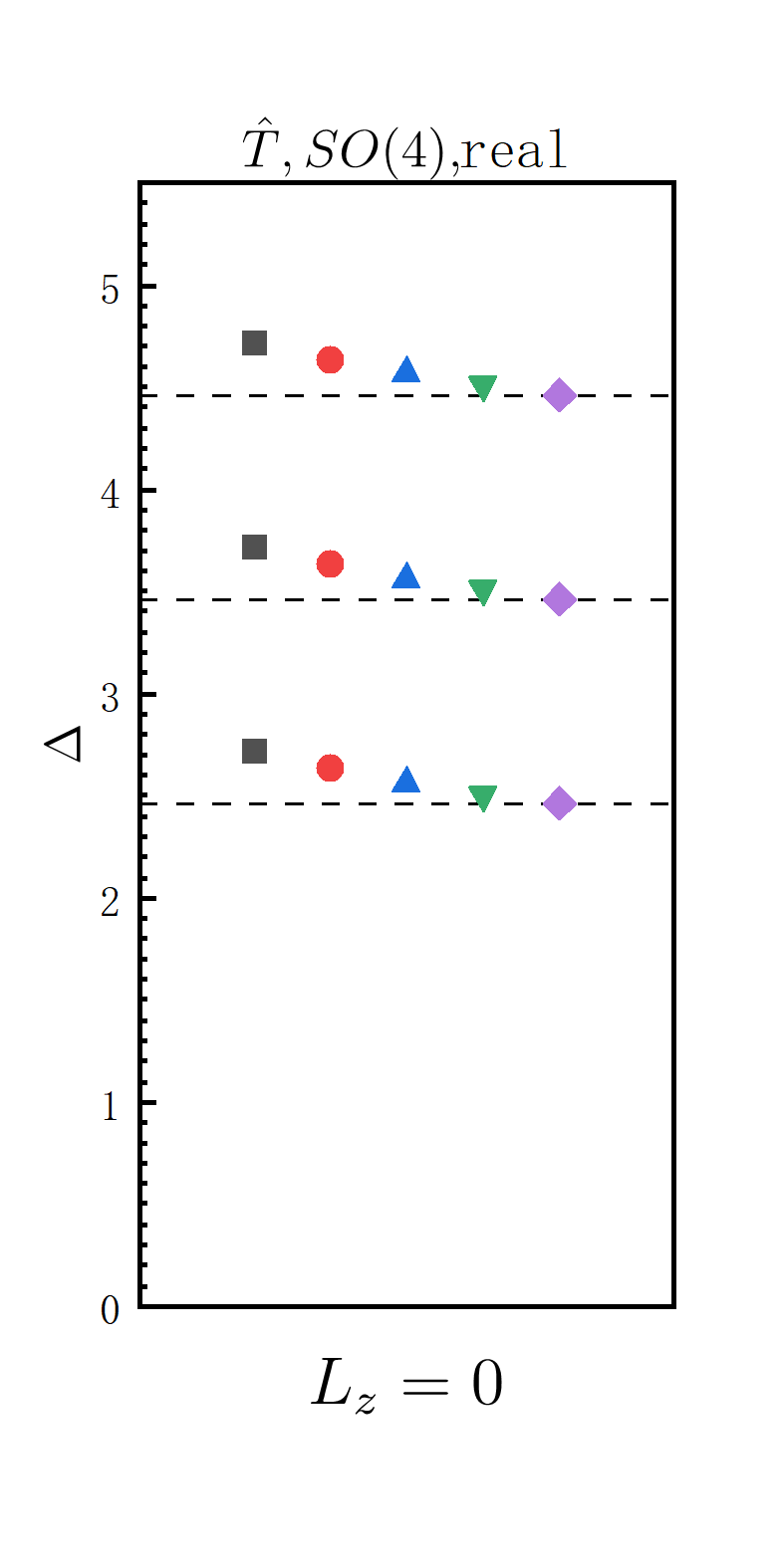}
\includegraphics[width=0.2\textwidth]{figure/SO4real/phia.png}
\includegraphics[width=0.2\textwidth]{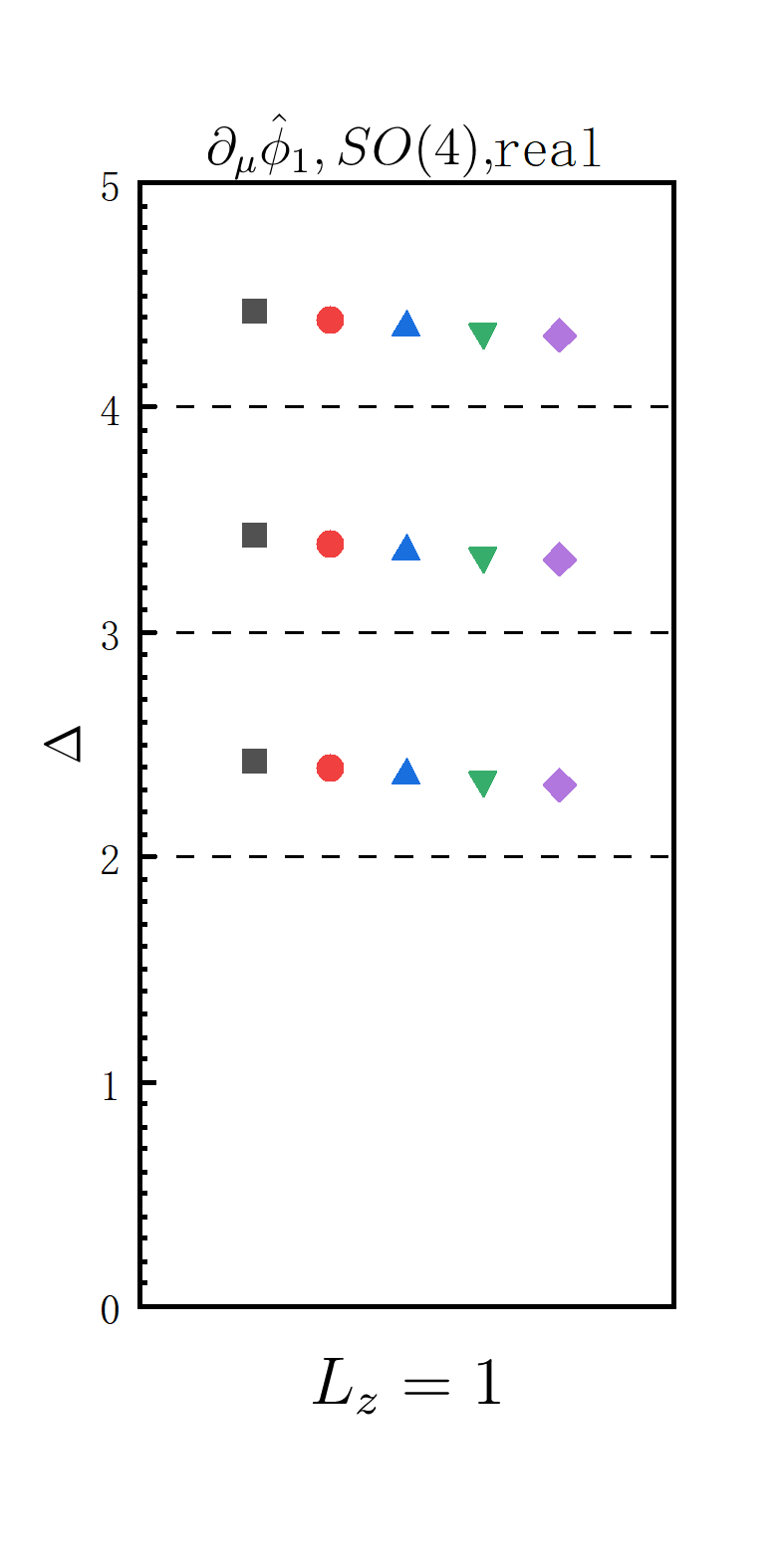}
\includegraphics[width=0.2\textwidth]{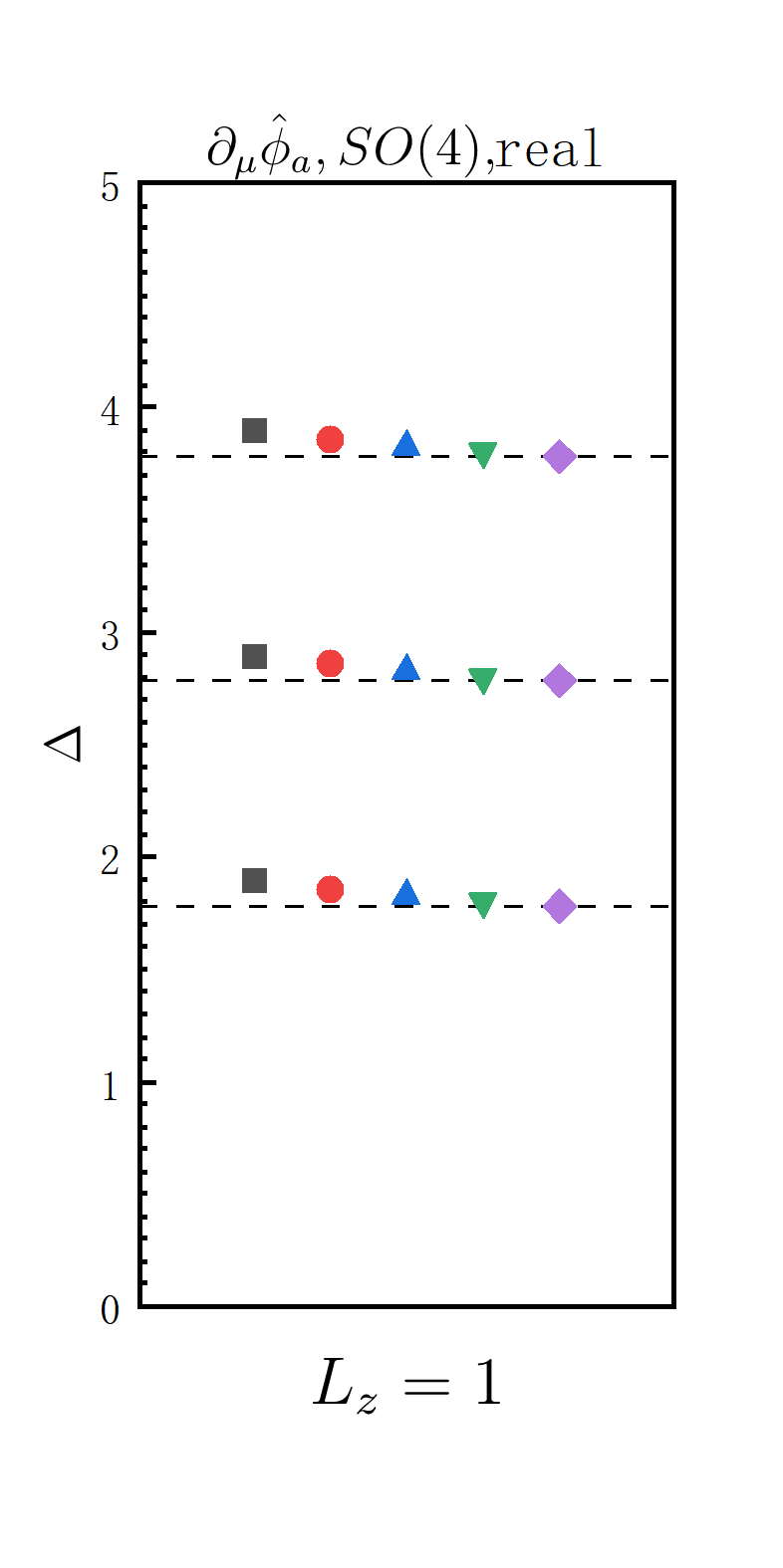}    
\includegraphics[width=0.2\textwidth]{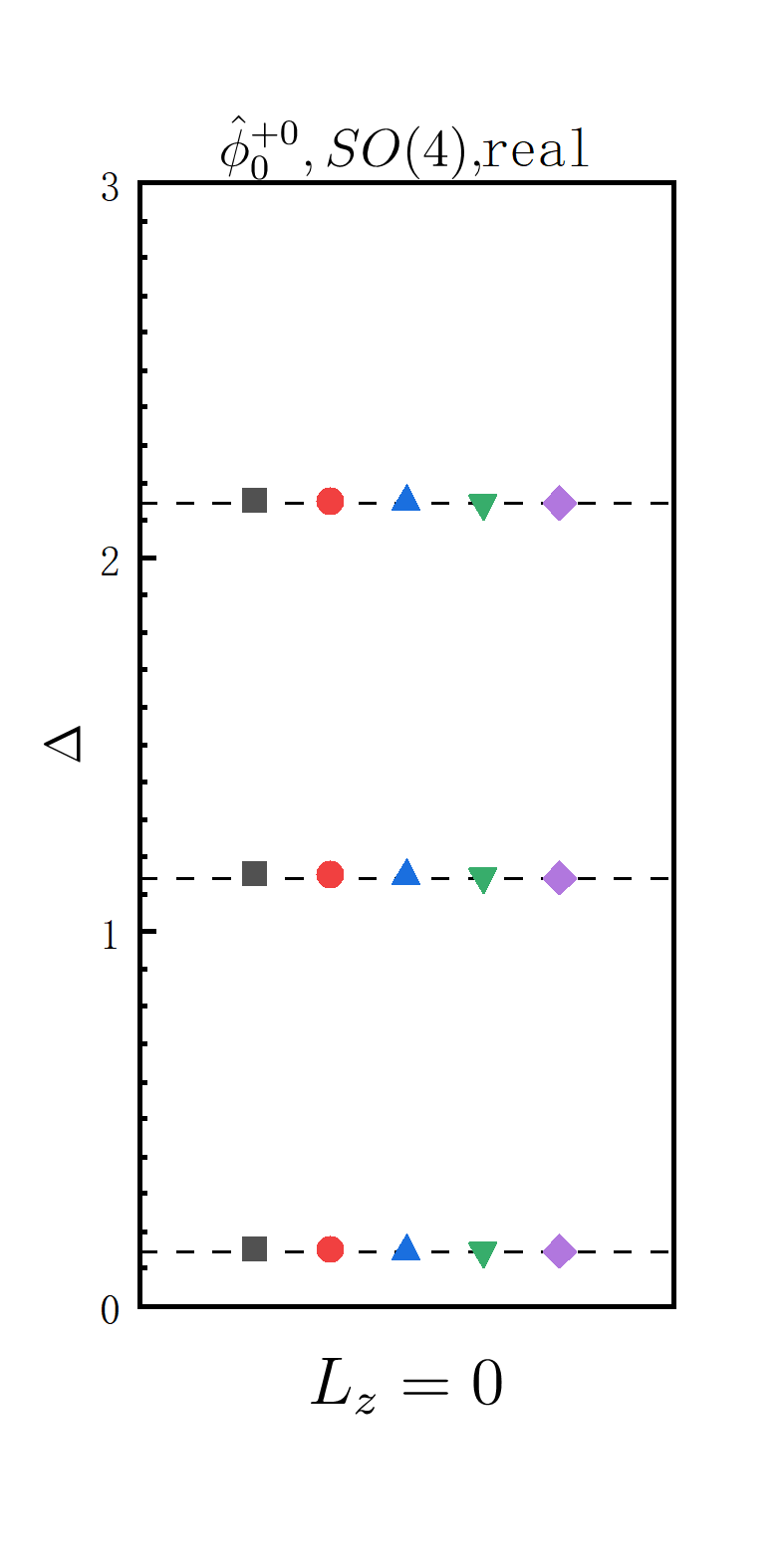} 
\includegraphics[width=0.2\textwidth]{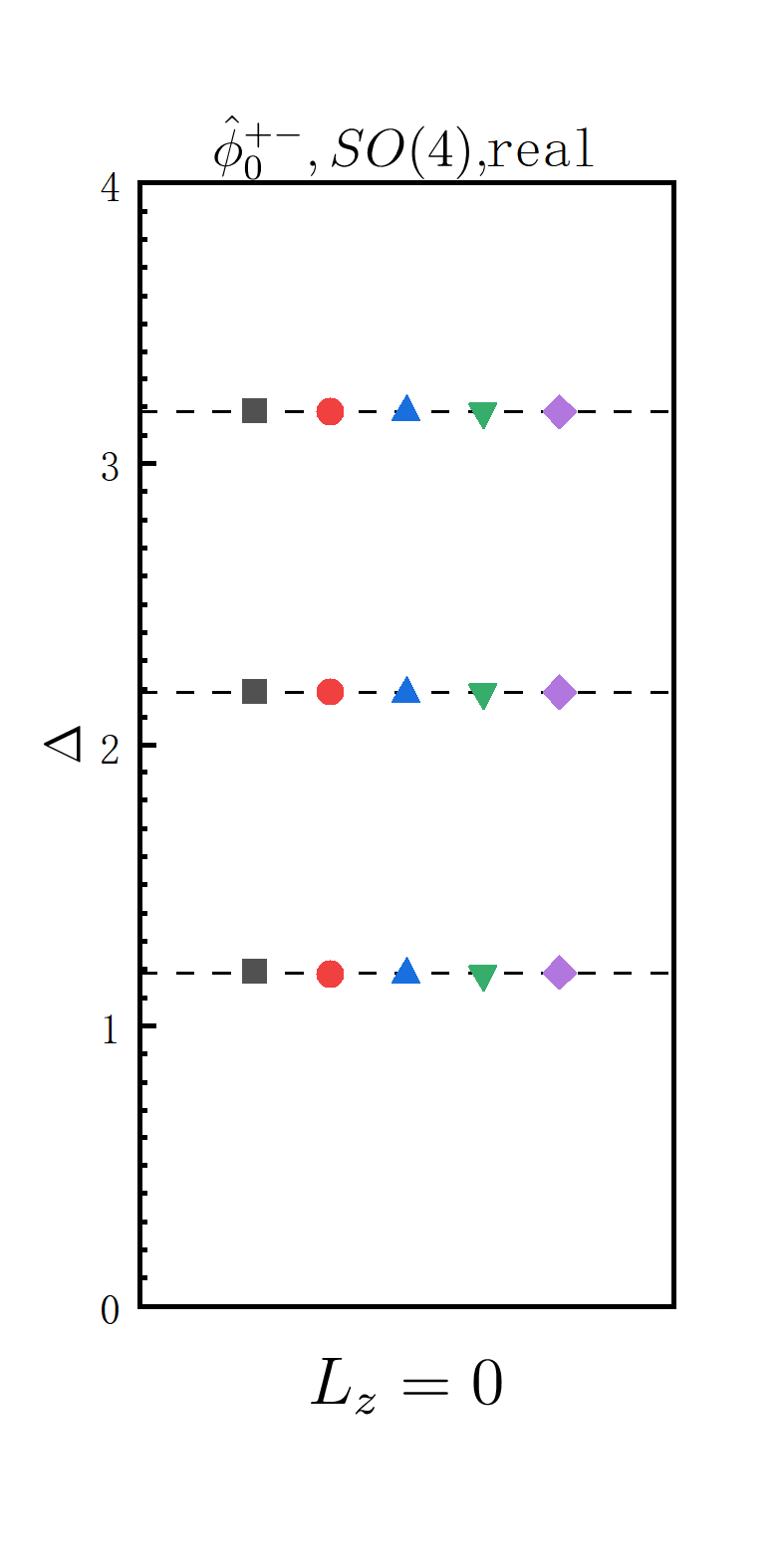}
    \caption{Line defect conformal multiplets. These results are produced at $h_d=1000$ in a pseudo-critical O(4) deconfined bulk. The ED result at the largest accessible system size ($N_o=10$) is set as a reference(gray horizontal dashed lines). The bulk parameters are set to be identical with Ref.\cite{ys_o4dqcp_l6vw-6z79}.}
    \label{fig:o4_dqcp_multi}
\end{figure*}

To demonstrate that the defect universality discussed in the main text is not specific to the SO(5) deconfined quantum critical point, we perform the same analysis for another candidate DQCP with O(4) symmetry. This provides an independent check on the robustness of the defect conformal fixed point. The bulk Hamiltonian is obtained as an O(4) deformation of the model studied in the main text and has been extensively investigated in previous literature as a candidate realizing a deconfined quantum critical point\cite{ys_o4dqcp_l6vw-6z79}.
On top of this bulk Hamiltonian, we introduce the same type of pinning-field defect by coupling local fields at the north and south poles of the fuzzy sphere,
\[
H_d = H_{O(4)}+h_Nn_3(\theta=0,\varphi)+h_Sn_3(\theta=\pi,\varphi)
\]
In the presence of the defect, the candidate primary operators at the O(4)-symmetric fixed point evolve in a manner closely analogous to the \(SO(5)\rightarrow O(4)\) symmetry reduction discussed in the main text. As illustrated in Fig.\ref{fig:o4_dqcp_multi}, at a representative coupling strength \(h_d=1000\), we extract the low-energy operator spectrum and identify a sequence of states that can be naturally organized into defect conformal multiplets. For each candidate primary operator, descendants appear with scaling dimensions separated by approximately integer intervals, as expected from defect conformal symmetry. 

Moreover, we identify several lowest primary operators in the spectrum through the state-operator correspondence. 
These results are summarized in Tab.\ref{tab:o4_dqcp_operator_dimensions}.
\begin{table*}[htbp!]
  \centering
  \caption{Scaling dimensions of several primary operators in the defect spectrum obtained from finite-size calculations. The bulk Hamiltonian is tuned to the critical point of the $O(4)$ deconfined quantum critical point. Here, $\text{Dim}$ and $C_2$ denote the dimension and the quadratic Casimir eigenvalue of the representations under the residual $SO(3)$ group.}
  
  \begin{tabular}{c|ccc|cccc|c}  
    \hline\hline
    Operator & $L_z$ & $\text{Dim}_{SO(3)}$  & $C_2^{\mathrm{SO(3)}}$ & $N_o=10$ & $N_o=9$ &$N_o=8$ &$N_o=7$ & Notes \\
    \hline 
    $\hat{\phi}_1$              & 0 & 1 & 0 & 1.8657 & 1.8890 & 1.9201 & 1.9589 &  \\
    $\hat{\phi}_a$              & 0 & 3 & 2 & 1.1551 & 1.1768 & 1.2035 & 1.2370 & tilt Op. \\
    $\hat{S}_-$                 & 0 & 1 & 0 & 2.8461 & 2.6872 & 2.7471 & 2.8194 &  \\
    $\hat{V}_a$                 & 0 & 3 & 2 & 2.8932 & 2.7360 & 2.7778 & 2.8354 &  \\
    $\hat{T}_{ab}$              & 0 & 5 & 6 & 2.4595 & 2.6045 & 2.6855 & 2.7934 &  \\
    $\partial_\mu \hat{\phi}_1$ & 1 & 1 & 0 & 2.3214 & 2.0321 & 2.0494 & 2.0757 & displacement Op. \\
    $\partial_\mu \hat{\phi}_a$ & 1 & 3 & 2 & 1.7808 & 1.7408 & 1.7617 & 1.7884 &  \\
    \hline
    $\hat{\phi}^{+-}$           & 0 & 1 & 0 & 1.1869 & 1.1833 & 1.1819 & 1.1839 & defect changing Op. \\
    $\hat{\phi}^{+0}$           & 0 & 1 & 0 & 0.1448 & 0.1463 & 0.1483 & 0.1510 & defect creation Op. \\
    \hline\hline 
\end{tabular}
\label{tab:o4_dqcp_operator_dimensions}
\end{table*}
Overall, the O(4) results closely parallel those obtained for the SO(5) DQCP, indicating that the existence of the nontrivial defect fixed point and its associated conformal structure are robust against changes in the bulk symmetry. This agreement further supports the universality of the defect critical behavior reported in the main text.

\bibliography{main.bib} 


\end{document}